%% file: main.tex
\documentclass[11pt]{article}
\usepackage{epsfig,changebar,amsmath,sint_new}
\usepackage[american]{babel}
\input mycommands
\begin{document}
  \input title.tex
  \input sect1.tex

  \input sect2.tex

  \input sect3.tex

  \input sect4.tex

  \input sect5.tex

  \input sect6.tex

  \input acknow.tex

  \input appa.tex

\input ref.tex
\end{document}

%% file: mycommands.tex
\def\rmO{{\rm O}}

\def\proof{\noindent{\sl Proof:}\kern0.6em}

\def\dual{\mathstrut^*\kern-0.1em}

\def\lvec#1{\setbox0=\hbox{$#1$}
    \setbox1=\hbox{$\scriptstyle\leftarrow$}
    #1\kern-\wd0\smash{
    \raise\ht0\hbox{$\raise1pt\hbox{$\scriptstyle\leftarrow$}$}}
    \kern-\wd1\kern\wd0}
\def\rvec#1{\setbox0=\hbox{$#1$}
    \setbox1=\hbox{$\scriptstyle\rightarrow$}
    #1\kern-\wd0\smash{
    \raise\ht0\hbox{$\raise1pt\hbox{$\scriptstyle\rightarrow$}$}}
    \kern-\wd1\kern\wd0}

\def\nabstar#1{\nabla\kern-0.5pt\smash{\raise 4.5pt\hbox{$\ast$}}
               \kern-4.5pt_{#1}}

\def\drvstar#1{\partial\kern-0.5pt\smash{\raise 4.5pt\hbox{$\ast$}}
               \kern-5.0pt_{#1}}

\def\MeV{{\rm MeV}}
\def\GeV{{\rm GeV}}

\def\fm{{\rm fm}}

\def\Nf{N_{\rm f}}
\def\psibar{\overline{\psi}}

\def\rhoprime{\rho\kern1pt'}
\def\rhobar{\bar{\rho}}
\def\rhobarprime{\rhobar\kern1pt'}
\def\rhobartilde{\kern2pt\tilde{\kern-2pt\rhobar}}
\def\rhobartildeprime{\kern2pt\tilde{\kern-2pt\rhobar}\kern1pt'}

\def\zetabar{\bar{\zeta}}
\def\zetaprime{\zeta\kern1pt'}
\def\zetabarprime{\zetabar\kern1pt'}
\def\zetar{\zeta_{\raise-1pt\hbox{\sixrm R}}}
\def\zetabarr{\zetabar_{\raise-1pt\hbox{\sixrm R}}}

\def\phiimpr{\phi_{\kern0.5pt\hbox{\sixrm I}}}

\def\diracstar#1#2{
    \setbox0=\hbox{$\gamma$}\setbox1=\hbox{$\gamma_{#1}$}
    \gamma_{#1}\kern-\wd1\kern\wd0
    \smash{\raise4.5pt\hbox{$\scriptstyle#2$}}}

\def\ba{b_{\rm A}}

\def\bp{b_{\rm P}}

\def\bm{b_{\rm m}}

\def\ca{c_{\rm A}}

\def\csw{c_{\rm sw}}

\def\f1{f_1}

\def\opprime#1{\setbox0=\hbox{${\cal O}$}\setbox1=\hbox{${\cal O}_{\rm #1}$}
    {\cal O}_{\rm #1}\kern-\wd1\kern\wd0
    \smash{\raise4.5pt\hbox{\kern1pt$\scriptstyle\prime$}}\kern1pt}

\def\ophatprime#1{\setbox0=\hbox{$\widehat{\cal O}$}
    \setbox1=\hbox{$\widehat{\cal O}_{\rm #1}$}
    \widehat{\cal O}_{\rm #1}\kern-\wd1\kern\wd0
    \smash{\raise4.5pt\hbox{\kern1pt$\scriptstyle\prime$}}\kern1pt}

\def\bopprime#1{\setbox0=\hbox{${\cal O}$}\setbox1=\hbox{${\cal O}_{\rm #1}$}
    {\cal L}_{\rm #1}\kern-\wd1\kern\wd0
    \smash{\raise4.5pt\hbox{\kern1pt$\scriptstyle\prime$}}\kern1pt}

\def\blagprime#1{\setbox0=\hbox{${\cal B}$}\setbox1=\hbox{${\cal B}_{#1}$}
    {\cal B}_{#1}\kern-\wd1\kern\wd0
    \smash{\raise5.2pt\hbox{\kern1pt$\scriptstyle\prime$}}\kern1pt}

\def\mq{m_{\rm q}}

\def\mc{m_{\rm c}}

\def\msbar{{\rm \overline{MS\kern-0.05em}\kern0.05em}}
\def\MSbar{{\rm \overline{MS\kern-0.05em}\kern0.05em}}

\def\bfy{\bf y}
\def\bfz{\bf z}

\def\Ms{M_{\rm s}}
\def\Mc{M_{\rm c}}
\def\mc{m_{\rm c}}

\def\mqc{m_{\rm q,c}}
\def\msc{m_{\rm sc}}

\def\mcmc{\overline{m}^\MSbar_{\rm c}(\overline{m}_{\rm c})}

%% file: title.tex
\begin{titlepage}
\begin{flushright}
   CERN-TH/2002-244 \\
   HU-EP-02/40\\
   FTUAM-02-22 \\
   September 2002
\end{flushright}
\vskip 1 cm
\begin{center}
  {\Large\bf  A precise determination of the charm quark's mass\\[1ex]
    in quenched QCD}
\end{center}
\vskip 1 cm
\begin{figure}[h]
\begin{center}
\epsfig{figure=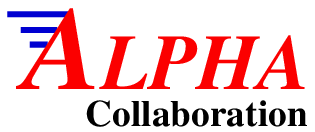} 
\end{center}
\end{figure}
\begin{center}
{\large Juri Rolf$^{\scriptscriptstyle a}$ and
Stefan Sint$^{\scriptscriptstyle b,c}$}
\vskip 2.3ex
\begin{itemize}
\item[$^{\scriptstyle a}$] Institut f\"ur Physik, Humboldt-Universit\"at zu 
Berlin,
Invalidenstr. 110, D-10115 Berlin, Germany
\item[$^{\scriptstyle b}$] 
CERN, Theory Division, CH-1211 Geneva 23, Switzerland
\item[$^{\scriptstyle c}$] 
Departamento de F\'{\i}sica Te\'orica, Universidad Aut\'onoma de Madrid,
28049 Cantoblanco (Madrid), Spain
\end{itemize}
\vskip 1.5cm
{\bf Abstract}
\vskip 0.7ex
\end{center}
We present a lattice determination of the charm quark's mass, 
using the mass of the $D_s$ meson as experimental
input. All errors are under control with the exception
of the quenched approximation. Setting the scale with $F_K=160\,\MeV$,
our final result for the renormalization group invariant (RGI)
quark mass is $M_{\rm c} = 1.654(45)\,\GeV$, which translates
to $\mcmc =1.301(34)\,\GeV$ for
the running mass in the $\MSbar$ scheme. 
A 6 percent increase of the RGI quark mass
is observed when the scale is set by the nucleon mass.
This is a typical quenched scale ambiguity, which 
is reduced to about 3 percent for $\mcmc$, and
to 4 percent for the mass ratio $\Mc/\Ms$. 
In contrast, the mass splitting 
$m_{D_s^\ast}-m_{D_s^{}}$ changes from $117(11)\,\MeV$
to $94(11)\,\MeV$, which is significantly smaller than the 
experimental value of $144\,\MeV$.
\vfill
\eject

\vfill

\eject

\end{titlepage}

%% file: sect1.tex
\section{Introduction}

The charm quark mass is among the fundamental parameters
of the Standard Model, and its determination
from experimental data is of general interest.
In practice it is important to improve on the
current precision: the Particle Data Group in its latest 
edition~\cite{PDG_review} gives the range 
\begin{equation}
       1.0 \leq \mcmc \leq 1.4\,\GeV,
\end{equation}
thereby increasing the previously quoted uncertainty~\cite{Groom:2000in}
by a factor two. The relatively large uncertainty in the 
charm quark mass has been identified 
as the dominant error in estimates of the fine structure constant
at high energies~\cite{Jegerlehner:2001ca}, and in phenomenological 
estimates of certain $B$-decay rates~\cite{Gambino:2001ew,Hurth:2001et}. 

A determination of quark masses from experiment is necessarily
indirect, as quarks are not observed as free particles.
To establish the connection one needs control over 
the strong interactions at the non-perturbative level. 
We distinguish two classes of approaches:
using some variant of the sum rule technique, 
one relies on perturbation theory and assumptions 
such as quark-hadron duality, which allow to connect 
to quantities derived from experiment~\cite{Martin:2000dd-Penarrocha:2001ig}.
While the quoted error especially in~\cite{Kuhn:2001dm} is rather small, 
it seems fair to say that a reliable assessment of the systematic
errors is difficult and may not always be possible.
 
The lattice approach, on the other hand, is non-perturbative in nature
and allows to directly compute hadronic observables 
at fixed lattice spacing and bare quark masses.
One may then determine the bare quark masses for which 
the experimental input is matched. In order to take the continuum 
limit one must substitute bare by renormalized masses, 
which are kept fixed as
the lattice spacing is varied. It should be emphasized that 
consistency requires the renormalization procedure 
to be non-perturbative, too.
In recent years there has been significant progress in non-perturbative
renormalization techniques~\cite{Sint:2000vc}. This has led
to the determination of the $\Lambda$-parameter
and the strange quark mass from low-energy hadronic observables, 
with no uncontrolled systematic errors apart from the quenched 
approximation~\cite{Capitani:1999mq,Garden:2000fg}.

In this paper we present a lattice computation of the charm
quark mass, where we use the same techniques as for 
the strange quark mass~\cite{Garden:2000fg}. In particular we take 
over the results for the non-perturbative quark mass 
renormalization of ref.~\cite{Capitani:1999mq}, as the renormalization
constant is quark mass and hence flavour independent.
In addition, all counterterms which are needed to 
remove the leading lattice artefacts in renormalized quark masses 
have been determined non-perturbatively in the relevant 
range of parameters~\cite{Guagnelli:2000jw,Bhattacharya:2000pn}. 
This is an important ingredient, as the charm quark mass is not very small
compared to the inverse lattice spacing, so that
cutoff effects can be large. Nevertheless, as will be demonstrated
below, controlled continuum extrapolations appear to be 
possible once the  leading O($a$) artefacts have been cancelled. 
For this reason our final result for the charm quark mass is much more 
precise than previous lattice 
estimates~\cite{Allton:1994ae-Becirevic:2001yh}. 
In particular, the dominant remaining uncertainty stems from the use of 
the quenched approximation, and further progress will require 
the inclusion of sea quark effects. 

The paper is organized as follows: In section~2
we explain our general strategy and the assumptions made. 
We then describe the technical framework of our calculation and 
collect the formulae needed for the quark mass 
renormalization and O($a$) improvement~(sect.3).
Section~4 contains some details of the numerical simulations 
and a discussion of systematic errors.
Continuum extrapolations and results for the charm quark mass, the charm
to strange quark mass ratio and the hyperfine splitting of 
$D$ mesons are discussed in sect.~5, and we conclude in sect.~6.

%% file: sect2.tex
\section{Strategy}

\subsection{Experimental input and basic assumptions}

It is generally assumed that the experimentally observed
light hadron spectrum can be accounted for by considering 
the physics of strong and electromagnetic 
interactions between up, down, strange and charm quarks.
Furthermore, due to the smallness of the fine structure
constant,
electromagnetic effects are likely to be small. 
Experimental results for the light
hadron spectrum can thus directly be compared to results of 
a QCD calculation, possibly after a small correction for  
electromagnetic effects.

Numerical simulations of lattice QCD offer the possibility
to compute the hadronic spectrum with an accuracy  
of typically a few per cent. One may then turn the
tables and use the precise experimental data 
to determine the free parameters
of QCD, i.e.~the gauge coupling and the quark masses.
For instance, assuming isospin symmetry, $M_{\rm u}=M_{\rm d}$, 
the four remaining parameters can be determined by
matching the kaon's decay constant and mass, 
\begin{equation}
    F_K = 160\,\MeV,\qquad m_K= 495\,\MeV,
\label{mK}
\end{equation}
the mass of the $D_s$ meson,
\begin{equation}
     m_{D_s} = 1969\,\MeV,
\end{equation}
and by taking the quark mass ratio,
\begin{equation}
   \Ms/\hat{M} = 24.4 \pm 1.5,
   \qquad \hat{M}=\tfrac12(M_{\rm u}+M_{\rm d}),
 \label{mass_ratio}
\end{equation}
as determined in Chiral Perturbation Theory~\cite{Leutwyler:1996qg}.
This latter input might be traded e.g.~for the pion mass. 
However, lattice simulations are typically 
carried out for quarks not very much lighter than the strange quark,
so that a chiral extrapolation becomes necessary. As the ansatz
for such an extrapolation is usually guided by chiral perturbation theory, 
the input of the pion mass is in practice not very different 
from the direct use of eq.~(\ref{mass_ratio}).
A similar problem occurs if one attempts to determine the kaon
mass and its decay constant at the physical values of the quark
masses. Chiral perturbation theory predicts a very
weak dependence upon the difference of the valence quark 
masses~\cite{Gasser:1983yg,Bijnens:1994qh},
and this has been verified numerically to some extent~\cite{Garden:2000fg}.
In practice one then computes with mass degenerate quarks
the sum of which matches the sum of the physical light and strange quark
masses. Once again, one relies on the assumption that
chiral perturbation theory provides a reliable description
of the quark mass dependence beyond the range actually covered by
the simulations.

\subsection{Electromagnetic effects}

Electromagnetic effects on hadronic observables involving up, down
and strange quarks can be computed in Chiral Perturbation
Theory. For instance, the result $m_K=495\,\MeV$~(\ref{mK}) is the 
(isospin averaged) experimental result after a subtraction of 
an estimate for the electromagnetic mass shift~\cite{Garden:2000fg}. 

For the charm quark mass determination we
would like to estimate electromagnetic effects on
$D$-mesons masses, where standard Chiral Perturbation Theory
does not apply\footnote{see, however, refs.~\cite{Wise:hn,Burdman:gh} for an
extension of Chiral Perturbation Theory to mesons containing a heavy quark.}. 
We propose a phenomenological estimate and shall assume
that all pseudoscalar $D$ meson masses can be parameterized in the form
\begin{equation}
 m_{\rm PS} = A M_{\rm light}  + B q + C,
\end{equation}
where $M_{\rm light}$ is the light valence quark mass and $q$ 
distinguishes electrically charged ($q=1$) 
and neutral ($q=0$) mesons. Note that a linear dependence on the 
light valence quark mass is indeed observed in
lattice simulations, at least for some range of light valence
quark masses~\cite{Bowler:2000xw}. 
Using this ansatz, we first note that $C$ cancels
in the mass differences, 
\begin{eqnarray}
   m_{D_s^\pm}-m_{D_{}^\pm} &=& A(M_{\rm s}-M_{\rm d}),\\
   m_{D_{}^\pm}-m_{D_{}^0} &=&  A(M_{\rm d}-M_{\rm u}) + B.
\end{eqnarray}
We may then eliminate $A$ and solve for the electromagnetic mass shift $B$,
\begin{equation}
   B =  m_{D_{}^\pm}-m_{D_{}^0} - 
  \dfrac{M_{\rm d}-M_{\rm u}}{M_{\rm s}-M_{\rm d}}
  \left(m_{D_s^\pm} - m_{D_{}^\pm}\right).
\end{equation}
Taking the mass ratio from Chiral 
Perturbation Theory~\cite{Leutwyler:1996qg},
\begin{equation}
 \dfrac{M_{\rm d}-M_{\rm u}}{M_{\rm s}-M_{\rm d}} = (40.3\pm 3.2)^{-1},
\end{equation}
and the experimental data for the mass splittings~\cite{Groom:2000in},
\begin{equation}
  m_{D_s^\pm}-m_{D_{}^\pm} = 99.2\pm 0.5\,\MeV, \qquad
  m_{D_{}^\pm}-m_{D_{}^0}= 4.79\pm 0.10\,\MeV,
\end{equation}
we arrive at the estimate
\begin{equation}
   B \approx 2.3 \,\MeV.
\end{equation}
Similar considerations can be made for the vector mesons,
where this effect is even smaller, 
and we conclude that electromagnetic effects cause 
$D$-meson mass shifts of a few MeV at most. 
This is below the accuracy currently reached by lattice computations,
so that we will neglect electromagnetic effects in the following.

Although there is little doubt that electromagnetic mass shifts
are indeed negligible we mention that 
this could also be checked  explicitly, 
by including an additional U(1) gauge field
in the numerical simulations~\cite{Duncan:1996xy}. 

\subsection{Quenched results for the $\Lambda_{\MSbar}$ and $M_{\rm s}$}

In the quenched approximation the programme outlined above has
already been completed except for the charm quark mass, which
is the topic of this work. Instead of quoting the value of a 
running coupling in some renormalization scheme at a reference
scale, it is customary to quote the $\Lambda$ parameter
in the $\MSbar$ scheme\footnote{
Note that $\Lambda_{\MSbar}$ can be 
defined beyond perturbation theory, owing to the fact that  
the {\em exact} relation between $\Lambda$-parameters of different schemes 
is determined by the one-loop perturbative relation between the respective couplings. 
Hence, $\Lambda_{\MSbar}$ can be defined indirectly, by referring 
to the $\Lambda$-parameter of a non-perturbatively defined
renormalization scheme.},
viz.~\cite{Capitani:1999mq}
\begin{equation}
      \Lambda_{\MSbar}^{(0)}   = 238(19)\,\MeV. 
\end{equation}
Furthermore, the result for the sum of the 
renormalization group invariant average
light and strange quark mass is~\cite{Garden:2000fg}
\begin{equation}
      \hat{M} + M_{\rm s}  = 143(5) \,\MeV.
\end{equation}
Using the quark mass ratio~(\ref{mass_ratio}),
the latter result translates to 
\begin{equation}
   M_{\rm s}= 138(6)\,\MeV\quad\Rightarrow\quad
   \overline{m}_{\rm s}^{\MSbar}(2\,\GeV) = 97(4)\,\MeV,
\end{equation}
for the running mass in the $\MSbar$ scheme of dimensional regularization
at the scale $2\,\GeV$. It should be emphasized that these results
have been obtained in the continuum limit, by carefully disentangling 
renormalization and cutoff effects. Thus, the only 
uncontrolled error arises from
the use of the quenched approximation.

\subsection{The scale $r_0$ and the quenched scale ambiguity}

The attempt to describe the real world with data obtained
within the quenched approximation is bound to fail at some point.
In particular, the attribution of physical units to 
quenched results is ambiguous.
To avoid this ambiguity we will quote results in units
of the low-energy scale $r_0$, which is derived from the force 
between static color sources~\cite{Sommer:1993ce}.
This has mostly technical advantages:
first, $r_0/a$  has been determined very precisely 
over a wide range of cutoff values~\cite{Guagnelli:1998ud-Necco:2001xg}, 
so that scaling studies are naturally carried out using this scale. 
Second, in the absence of dynamical quarks, $r_0$ 
is only affected by cutoff effects of order $a^2$~\cite{Necco:2001xg}.
The relation to other hadronic scales
in the quenched approximation is known~\cite{Garden:2000fg},
\begin{equation}
   r_0 F_K   =0.415(9), \qquad  r_0 m^{}_N \approx 2.6.
\end{equation}
This illustrates the inconsistency of the quenched approximation,
as the quenched result for  $F_K/m_N$ differs by 10 per cent from 
its experimental value. This can be viewed as a 
scale ambiguity: setting the scale
with $F_K=160\, \MeV$ is roughly equivalent 
to the choice $r_0=0.5\,\fm$, while
$m^{}_N=938\,\MeV$ corresponds to $r_0=0.55\,\fm$.
We will later investigate the effect of this ``quenched
scale ambiguity'', in order to get a first idea of the
size of typical quenching errors.

The results in the previous subsection have been
obtained using $r_0=0.5\,\fm$. In units of $r_0$ 
we have
\begin{equation}
   r_0 \Lambda_{\MSbar}^{(0)} = 0.602(48),\qquad
   r_0 M_{\rm s} =  0.348(13).
 \label{r0Lambda}
\end{equation}

\subsection{Strategy to compute the charm quark's mass}

At fixed cutoff, standard  lattice QCD techniques allow to compute 
meson masses for given bare mass parameters of the meson's valence quarks.
As for the strange quark mass, the results of~\cite{Garden:2000fg}
allow to set its bare mass parameter without further tuning.
To determine the bare charm quark mass it thus remains
to measure pseudoscalar meson masses for several bare mass parameters, 
and to interpolate the meson masses versus the bare charm quark mass
to the physical point where
\begin{equation}
     m_{D_s}=1969\,\MeV \quad\Rightarrow \quad r_0m_{D_s} = 4.99.
\end{equation}
Then, using known renormalization factors, one may obtain
the renormalization group invariant charm quark mass, $M_{\rm c}$,
in units of $r_0$ and at the given value of the cutoff $a^{-1}$.
Repeating the procedure for smaller lattice spacings $a$, one may
eventually extrapolate to the continuum. Additional
control of this extrapolation may be obtained by considering
alternative definitions of $r_0M_{\rm c}$, with different
cutoff effects but the same continuum limit.

In the following two sections we describe in some detail the technical
setup to achieve this goal. The reader not so much interested
in the technicalities may directly go to sect.~5, where the
continuum extrapolations and final results are discussed.

%% file: sect3.tex
\section{The technical framework}

For the practical implementation of the programme 
we use the framework of O($a$) improved lattice QCD
as described in ref.\cite{Luscher:1996sc}. For unexplained notation
we refer to this paper.

\subsection{SF correlation functions}

In order to extract hadron masses we use correlation functions 
derived from  the QCD Schr\"odinger 
functional (SF)~\cite{Luscher:1992an,Sint:1993un}. 
Hence, QCD is considered on a space-time manifold which is 
a hyper-cylinder of volume $L^3\times T$,
with the quantum fields satisfying periodic boundary conditions in space, and
Dirichlet boundary conditions at Euclidean times $x_0=0$ and $x_0=T$. 
The technical advantages for the computation of hadron properties 
have been demonstrated in~\cite{Guagnelli:1999zf}. 
One of the main points is the possibility
to define quark and antiquark boundary states 
$\zeta$ and $\zetabar$, which only transform
under spatially constant gauge transformations.
Boundary sources with quantum numbers of mesons are therefore
gauge invariant even if quark and antiquark fields
are localized at different points in space.

For the study of pseudoscalar and vector mesons, 
we define the axial current and density,
\begin{equation}
   A_\mu(x) = \psibar_i(x)\gamma_\mu\gamma_5\psi_j(x),\qquad
   P(x) = \psibar_i(x)\gamma_5\psi_j(x),
\end{equation} 
and the local vector current,
\begin{equation}
   V_\mu(x) = \psibar_i(x)\gamma_\mu\psi_j(x),
\end{equation} 
with flavour indices $i\ne j$. The correlation functions
\begin{eqnarray}
 f_{\rm A}(x_0) &=&  - \frac12 a^6 \sum_{\bfy,\bfz}
   \left\langle \bar\zeta_j(\bfy)\gamma_5\zeta_i(\bfz) A_0(x)\right\rangle,
 \label{fA} \\
 f_{\rm P}(x_0) &=&  - \frac12 a^6 \sum_{\bfy,\bfz}
   \left\langle \bar\zeta_j(\bfy)\gamma_5\zeta_i(\bfz) P(x)\right\rangle, \\
 k_{\rm V}(x_0) &=&  - \frac16 a^6 \sum_{\bfy,\bfz}
   \left\langle \bar\zeta_j(\bfy)\gamma_k\zeta_i(\bfz) V_k(x)\right\rangle, 
 \label{kV}
\end{eqnarray}
are then used to compute pseudoscalar and vector meson masses with
valence quark flavours $i$ and $j$. The notation and technology
has been described in detail in ref.~\cite{Guagnelli:1999zf}.
One studies the effective masses
\begin{equation}
    am_{\rm eff}(x_0+\tfrac12 a) =  \ln\left\{ f(x_0)/f(x_0+a)\right\},  
\end{equation}
where $f(x_0)$ stands for one of the above correlation functions.
For large time distances from the boundaries, i.e.~large
$x_0$ and $T-x_0$, one expects the ground state
of a given channel to dominate the correlation function,
which manifests itself in a plateau for the effective mass
versus $x_0$. 

\subsection{Definition of renormalized quark masses}

The renormalization of quark masses for Wilson type quarks is complicated
by the fact that all axial symmetries are
explicitly broken by the regularization. As a consequence, quark
mass renormalization is both additive and multiplicative,
the axial current requires a scale-independent renormalization,
and axial Ward identities, such as the PCAC relation are
violated by cutoff effects. There are various ways to define 
renormalized quark masses, which are  equivalent
in the continuum limit, but may differ at finite lattice spacing. 
This will later be used to to achieve a better control of 
the continuum extrapolations.

We start with the definition of a bare current quark mass 
using the PCAC relation, 
\begin{equation}
   m_{ij} = \dfrac{\tilde\partial_0 f_{\rm A}(x_0)}{2f_{\rm P}(x_0)},
  \label{PCACmass}
\end{equation}
where $\tilde\partial_\mu=\frac12(\partial^{}_\mu+\partial^\ast_\mu)$ 
denotes the symmetric lattice derivative in $\mu$-direction.
From the bare quark mass $m_{ij}$ we obtain the 
sum of the renormalized valence quark masses through
multiplicative renormalization of the axial current
and density in (\ref{PCACmass}), viz.
\begin{equation}
   m_{{\rm R},i}+m_{{\rm R},j} = 2 Z_{\rm A}Z_{\rm P}^{-1} m_{ij}.
\end{equation}
A single quark mass is obtained if the quark flavours $i$ and
$j$ are mass degenerate,
\begin{equation}
   m_{{\rm R},i} = Z_{\rm A}Z_{\rm P}^{-1} m_{i}.
\end{equation}

Another starting point is provided by the hopping parameter in the
Wilson quark action. One defines a bare subtracted
quark mass, 
\begin{equation}
    m_{{\rm q},i}=\frac12(\kappa_i^{-1}-\kappa_{\rm critical}^{-1}),
 \label{mq_def}
\end{equation}
where $\kappa_{\rm critical}$ is 
the value of $\kappa$ at which chiral symmetry is restored. 
This subtracted bare quark mass is then  multiplicatively renormalized,
\begin{equation}
    m_{{\rm R},i} = Z_{\rm m} m_{{\rm q},i}.
\end{equation}
In what follows, we will always refer to the same 
renormalization scheme for the quark mass.
This is achieved by first defining the renormalized axial current 
and density, and by setting
\begin{equation}
   Z_{\rm m} = Z Z_{\rm A}/Z_{\rm P},
\end{equation}
where $Z$ is the scale independent ratio between the
bare quark masses,
\begin{equation}
    m_{i} = Z m_{{\rm q},i}.
\end{equation}
Furthermore, we restrict attention to quark mass independent 
renormalization schemes.
The SF scheme of ref.~\cite{Sint:1998iq} has this property, and it
has the additional virtue of being defined beyond perturbation theory.
Its  relation to the renormalization group invariant (RGI) quark mass, 
\begin{equation}
   M_i = \lim_{\mu\rightarrow\infty} \overline{m}_i(\mu)
   \left[2b_0\bar{g}^2(\mu)\right]^{-d_0/2b_0},
\end{equation}
has been determined for $\Nf=0$~\cite{Capitani:1999mq}.
We recall that $b_0=(11-\frac23\Nf)/(4\pi)^2$ and $d_0=8/(4\pi)^2$
are the lowest order perturbative coefficients of the renormalization
group functions for gauge group SU(3). 
The running mass $\overline{m}(\mu)$ coincides
with $m_{\rm R}$ at some scale $\mu=\mu_0$ and is otherwise
determined by the RG equations.
The RGI quark mass is
non-perturbatively defined, it is scheme and
scale independent and hence a natural candidate for a fundamental
parameter of QCD. However, it is still customary to 
quote quark masses in the $\MSbar$ scheme of dimensional
regularization at some reference scale. Once $M$ is given,
the ratio $\overline{m}^{\MSbar}(\mu)/M$ can be computed 
by using the perturbative renormalization group functions, which are known
with 4-loop accuracy~\cite{vanRitbergen:1997va,Chetyrkin:1997dh}.

\subsection{O($a$) improvement}

In practice, cutoff effects with Wilson quarks can be rather 
large, and  it is advisable to cancel at least the
leading O($a$) artefacts in physical observables. 
One may distinguish cutoff effects which arise already 
in the chiral limit and those proportional
to the quark masses. To cancel the former, a single 
counterterm, the so-called Sheikholeslami-Wohlert 
term~\cite{Sheikholeslami:1985ij},
must be included in the action, with a coefficient $\csw$ which is
a function of the bare coupling only. For the improvement
of on-shell correlation function one also needs to improve
the operators. In our context, only the improvement
of the axial current is relevant,
\begin{equation}
   (A_{\rm I})_\mu = A_\mu + \ca a\tilde\partial_\mu P,
\end{equation}
with the improvement coefficient $\ca$.
Both coefficients are known non-perturbatively 
for $\Nf=0$ and bare couplings $g_0^2=6/\beta \leq 1$~\cite{Luscher:1996ug}.

O($a$) artefacts which are proportional to the quark mass $\mq$
can be dealt with perturbatively, as long as $a\mq$ is
small enough. In practice this may still be the case for the strange
quark, but in the case of charm the bare quark mass in lattice units
is typically in the range $0.25 - 0.5$. 
In the quenched approximation, the improvement coefficients needed
to compute O($a$) improved renormalized quark masses are
all known non-perturbatively. Renormalization and O($a$) improvement of
the subtracted bare quark mass is achieved separately for
each quark flavour through
\begin{equation}
   m_{\rm R} = Z_{\rm m} m_{\rm q}(1+b_{\rm m} am_{\rm q}).
\end{equation}
As for the PCAC quark mass, we first 
recall the form of the renormalized O($a$) improved axial
density and current,
\begin{eqnarray}
   (A_{\rm R})_\mu &=& Z_{\rm A}\left[
                      1+b_{\rm A}\tfrac12(am_{{\rm q},i}+am_{{\rm q},j})
                               \right](A_{\rm I})_\mu, \\
     P_{\rm R}     &=& Z_{\rm P}\left[
                      1+b_{\rm P}\tfrac12(am_{{\rm q},i}+am_{{\rm q},j})
                               \right] P.
\end{eqnarray}
For the quark mass improvement this means
\begin{equation}
  m_{{\rm R},i}+m_{{\rm R},j} = 2 Z_{\rm A}Z_{\rm P}^{-1}\left[
                      1+(b_{\rm A}-b_{\rm P})
                      \tfrac12(am_{{\rm q},i}+am_{{\rm q},j})
                                \right] m_{ij}.
\end{equation}
Here, it is implicitly assumed that the bare mass $m_{ij}$
is defined including the counterterm $\propto \ca$, i.e.
\begin{equation}
   m_{ij} = \dfrac{\tilde\partial_0 f_{\rm A}(x_0)
   +\ca a\partial^\ast_0\partial_0 
    f_{\rm P}(x_0)}{2f_{\rm P}(x_0)}.
\end{equation}
Besides this definition with standard lattice derivatives we
also use improved derivatives as in ref.~\cite{Guagnelli:2000jw}.
While the difference is formally an effect of O($a^2$) it 
may be important numerically.

The improvement coefficients $\bm$ and $\ba-\bp$ have been determined
for $\Nf=0$ in the relevant range of bare couplings~\cite{Guagnelli:2000jw}. 
This, together with the renormalization constants $Z$~\cite{Guagnelli:2000jw}, 
$Z_{\rm A}$~\cite{Luscher:1996jn} and $Z_{\rm P}$~\cite{Capitani:1999mq} 
allows us to define the renormalized O($a$) improved
charm quark mass in the SF scheme in various ways. 
One may then convert from the SF scheme to the RGI quark mass,
using the flavour independent ratio $M/\overline{m}(\mu_0)$
in the continuum limit~\cite{Capitani:1999mq}.
It is convenient to combine the factors relating the bare current quark
mass and the RGI masses into a single one,
\begin{equation}
  Z_M = \dfrac{M}{\overline{m}(\mu_0)}Z_{\rm A}Z_{\rm P}^{-1},
\end{equation}
which has been parameterized in~\cite{Capitani:1999mq} over some
range of bare couplings.
Up to O($a^2$) we then have
\begin{equation}
   M_i = Z_M \left[1+(\ba-\bp)am_{{\rm q},i})\right] m_i 
   = Z_M Z m_{{\rm q},i}(1+\bm am_{{\rm q},i}),
\end{equation}
and an analogous equation for $M_i+M_j$ in 
the mass non-degenerate case.

%% file: sect4.tex
\section{Numerical simulations}

The numerical simulations were carried out on machines of the APE100 and
APE1000 series using single precision arithmetic
and standard algorithms (see ref.~\cite{Luscher:1996ug} for a detailed
description and references).
Autocorrelation times were found to be small so that our measurements 
could be treated as statistically independent for all observables considered.

\subsection{Simulation parameters}

Our choice of simulation parameters is displayed in table~\ref{tab_param}.
The non-perturbatively determined coefficients $\csw$ and $\ca$
were taken from ref.~\cite{Luscher:1996ug}. 
We chose the same four values of $\beta$ as in~\cite{Garden:2000fg},
which correspond lattice spacings in the range 
$a=0.05 - 0.1\, {\rm fm}$. The conversion to physical 
units is done using the scale $r_0 = 0.5\,\fm$~\cite{Sommer:1993ce} 
and the parameterization of ref.~\cite{Guagnelli:1998ud},
\begin{equation}
   \ln({a}/{r_0}) = -1.6805 - 1.7139 (\beta-6) + 0.8155 (\beta-6)^2 
                    -0.6667 (\beta-6)^3.
\end{equation}
The lattice size $L/a$ varied between $16$ and $32$, 
such that the linear extent of the spatial volume
was around $L \simeq 1.5 \,\fm$ or larger. For spatial
volumes of this size we do not expect any sizeable 
finite volume effects on our pseudoscalar and vector masses.
This is supported by previous findings e.g.~in ref.~\cite{Garden:2000fg}, 
where the pseudoscalar meson mass was found to be safe 
against finite volume effects, provided its Compton wave length was 
about 5 times smaller than $L$. Similar statements are expected to hold 
for the vector mesons. In the time direction the lattice size 
varied between $40$ and $80$, such that the physical 
extent $T/r_0$ was kept constant to a good approximation. 
\begin{table}[htb]
 \centering
 \begin{tabular}{l c c c c r c r}
 \hline \\[-1.0ex]
 $\beta$ & $T/a$ & $L/a$ & $L/r_0$ & $T/r_0$ & $r_0/a\quad$ 
 & $N_{\rm meas}$ \\[1.0ex]
 \hline \\[-1.0ex]
  6.0 & 40 & 16 &  $2.98(1)$ & $7.45(3)$ & $5.368(22)$   & 350  \\
  6.1 & 48 & 24 &  $3.80(2)$ & $7.59(3)$ & $6.324(28)$   &  94  \\
  6.2 & 54 & 24 &  $3.26(2)$ & $7.34(4)$ & $7.360(35)$   & 159  \\
  6.45& 80 & 32 &  $3.06(2)$ & $7.65(4)$ & $10.458(58)$  & 123  \\[1.0ex]
 \hline
 \end{tabular}
 \caption[tab_param]{Simulation parameters,
 lattice sizes and number of independent measurements.
 \label{tab_param}}
\end{table}

Our estimates of the hopping parameter for the strange quark
are based on table~2 of ref.~\cite{Garden:2000fg}
where the quantity 
\begin{equation}
   \left[Z_M R/r_0\right]_{r_0^2m_{\rm PS}^2=1.5736} 
   = 1.5736\times r_0(M_{\rm s}+\hat{M}),
\end{equation}
is quoted for each value of $\beta$.
In order to isolate the strange quark mass we use the mass 
ratio from chiral perturbation theory~(\ref{mass_ratio}). 
Then we relate the RGI strange quark mass to the 
subtracted bare quark mass in lattice units,
\begin{equation}
  r_0 M_{\rm s} = (r_0/a) Z_M Z a m_{\rm q,s}(1+\bm am_{\rm q,s}). 
 \label{m_strange}
\end{equation}
where we use the coefficients  $Z$ and $\bm$ of 
ref.~\cite{Guagnelli:2000jw}.
Finally, to relate to the corresponding hopping parameter $\kappa_{\rm s}$
(\ref{mq_def}), we use estimates of $\kappa_{\rm critical}$ at 
$\beta=6.0,6.2$~\cite{Luscher:1996ug},
and interpolations at $\beta=6.1,6.45$~\cite{Hartmut} 
(cf.~table~\ref{tab_renorm}). In this way the hopping parameter
for the strange quark could be fixed prior to the simulations.
As usual, there is some uncertainty associated with the choice
of $\kappa_{\rm critical}$. In order to assess the sensitivity
of our observables to the precise choice of the strange quark mass,
we also used  a second value of $\kappa_{\rm s}$ for all but
our largest $\beta$-value. 
Finally, we chose 3 $\kappa$-values around the expected
physical charm quark mass. The numerical values for
all simulated hopping parameters are displayed in table~\ref{tab_sim1}.
\input tab_sim1.tex

\subsection{$D$-meson masses}

From the correlation functions~(\ref{fA}--\ref{kV}) we 
determined the corresponding effective masses $m_{\rm X}(x_0)$
in the pseudo-scalar (${\rm X = PS}$) and vector channel
(${\rm X = V}$). In addition we considered the effective
mass $m_{\rm S}$ associated with the ratio 
$f(x_0) = k_{\rm V}(x_0)/f_{\rm P}(x_0)$,
which directly yields the mass splitting between the vector 
and the pseudoscalar channel. 
For all parameter choices these effective masses exhibit a plateau.
Deviations from the plateau value at
small $x_0$ or small $T-x_0$ are expected, 
due to the contribution of excited states. 
Furthermore, since the masses in our work 
are quite large, the correlation functions decay rapidly 
and one might expect problems 
with rounding errors at the larger values of $x_0$.

\begin{figure}[h]
\label{fig_rounding}
\centering
\epsfig{file=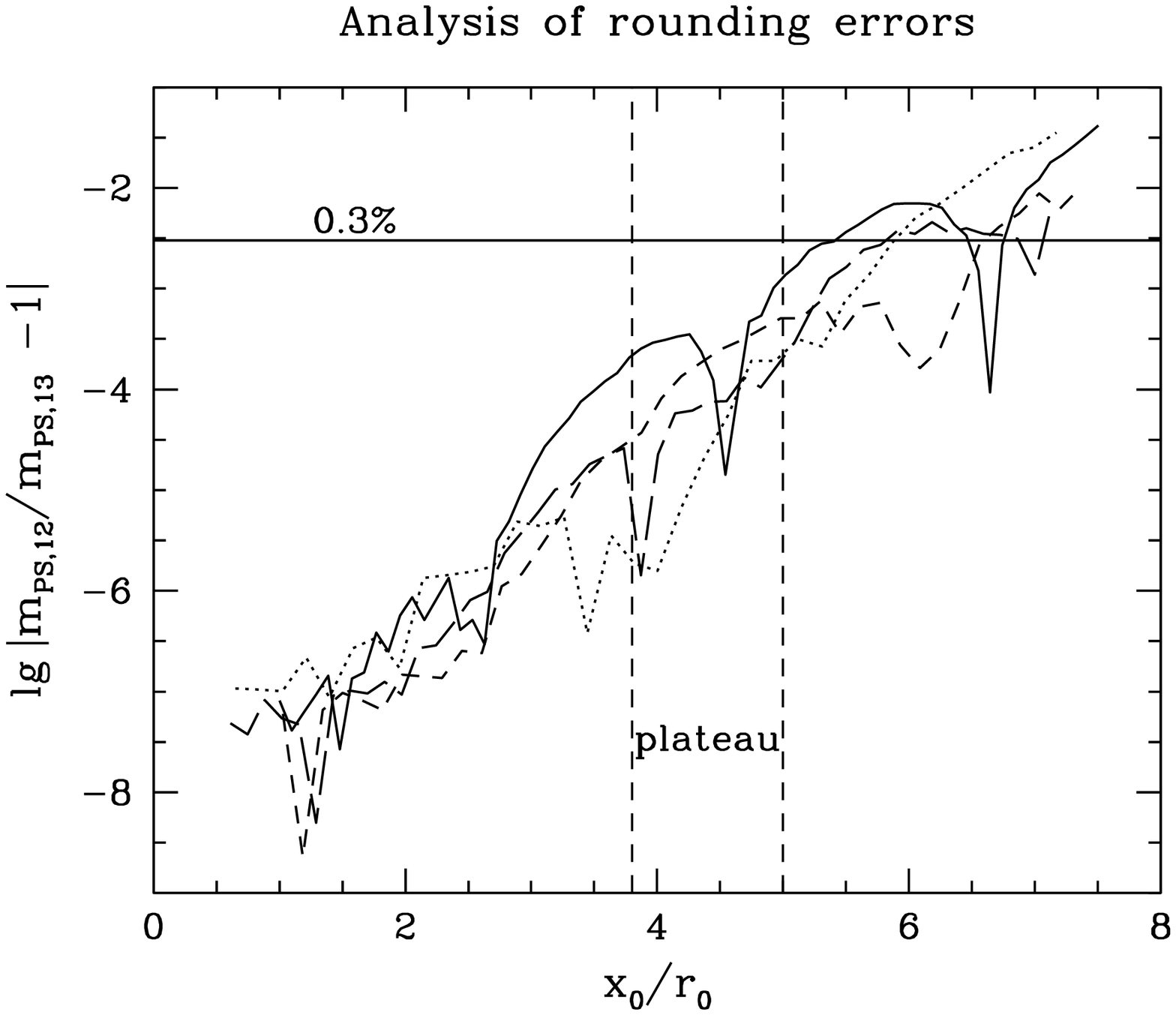, width=.8\linewidth,
        bbllx=18,bblly=184,bburx=572,bbury=658}
\caption{Illustration of the rounding error problem: 
The plot shows, for all 4 $\beta$-values, 
the logarithm of the relative differences between
two effective masses which differ by the solver precision used (cf.~text). 
The curves are obtained as averages over O(10) configurations
using the $\kappa_{\rm c}$-value closest to the physical point.
The chosen plateau region is indicated by the vertical lines.
}
\end{figure}

\subsubsection{Rounding errors}

To estimate rounding errors, we varied the precision of the 
quark propagator computation. In our production runs 
the Wilson-Dirac equation was solved with a 
squared relative precision $\epsilon^2 = 10^{-14}$.
In addition, at each $\beta$ value we chose
ten independent gauge field configurations, and repeated
the calculation with $\epsilon^2 = 10^{-12}$ and $\epsilon^2 = 10^{-13}$.
Indicating the solver precision by a superscript, we find that
the ratio of effective masses 
${{m_{\rm X}}^{(12)}}/{{m_{\rm X}}^{(13)}}-1$ grows
roughly exponentially as a function of the time coordinate. 
We performed an exponential fit to this ratio of the form 
$r \exp(x_0 R)$, which was then 
used to estimate rounding errors on the effective masses.
Requiring that the relative error on the masses not exceed
a certain value (typically a fraction of a percent)
then leads to an upper limit for the 
range of the plateau. However, note that our method is likely 
to overestimate this systematic effect, as the limit 
of single precision is only reached for
solver residuals between $\epsilon^2 = 10^{-13}$ and 
$\epsilon^2 = 10^{-14}$.
\par\noindent
\begin{figure}[t]
\begin{minipage}[h]{\linewidth}
\epsfig{file=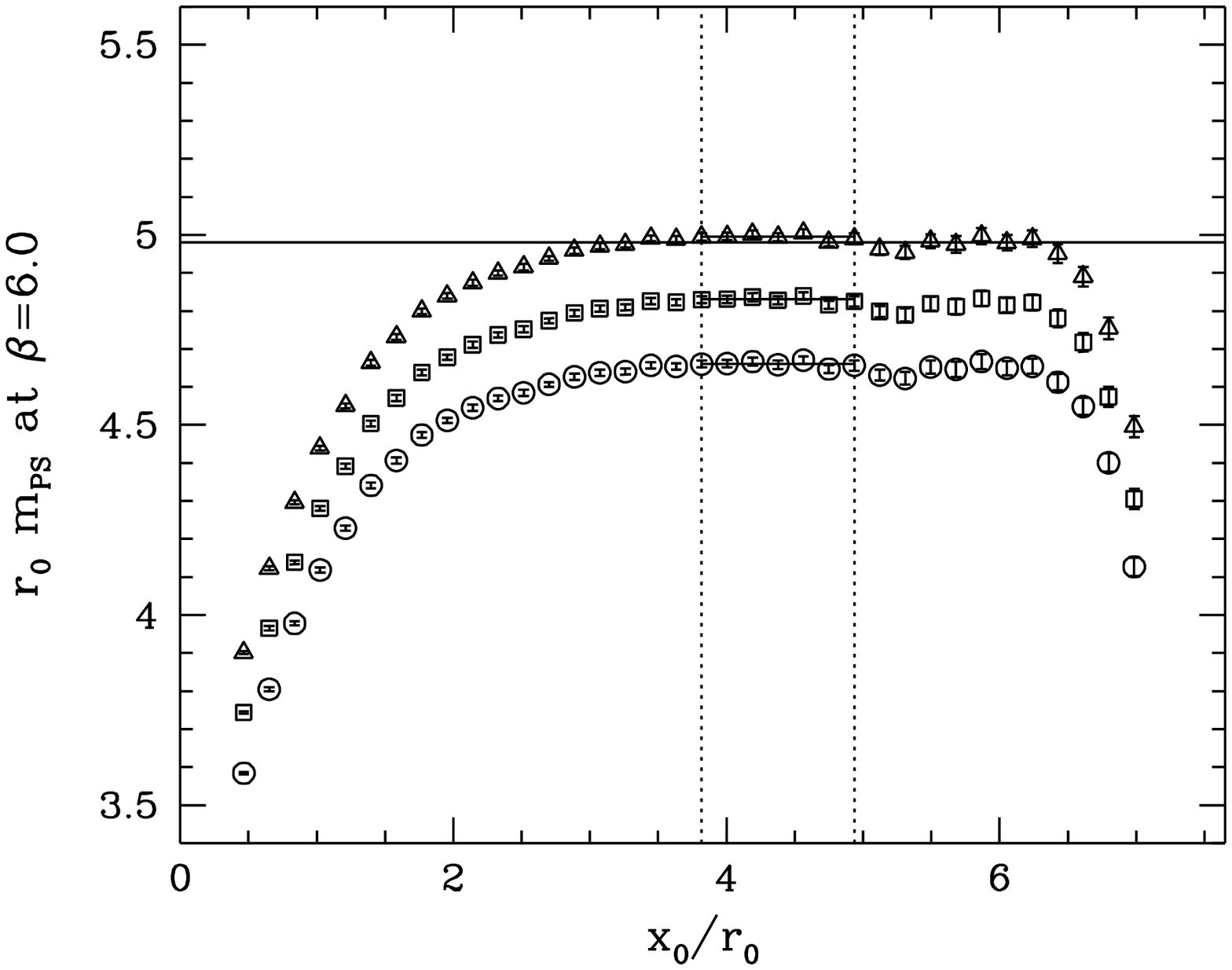, width=.48\linewidth,
        bbllx=18,bblly=144,bburx=572,bbury=518}
\epsfig{file=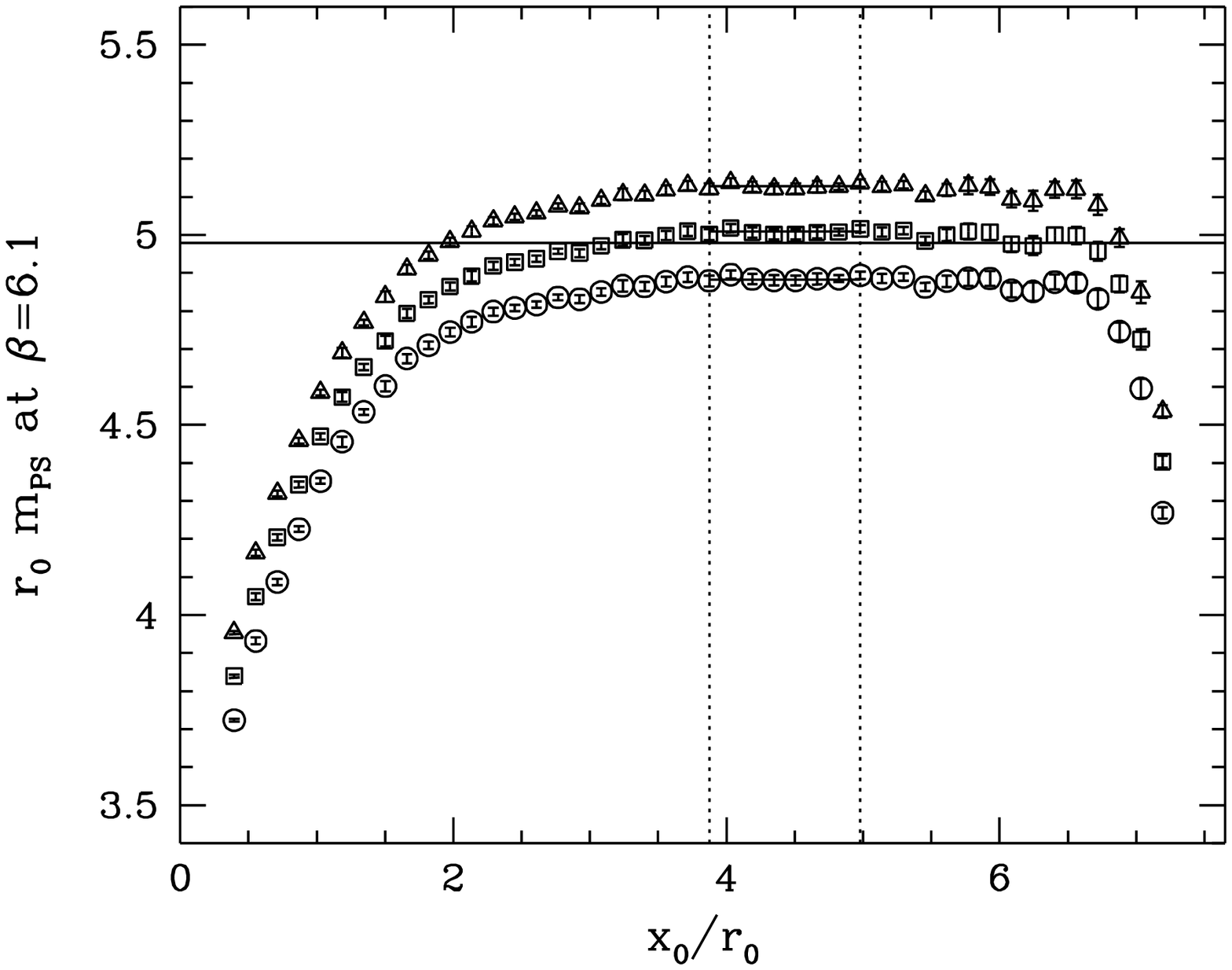, width=.48\linewidth,
        bbllx=18,bblly=144,bburx=572,bbury=518}\hfill\\
\epsfig{file=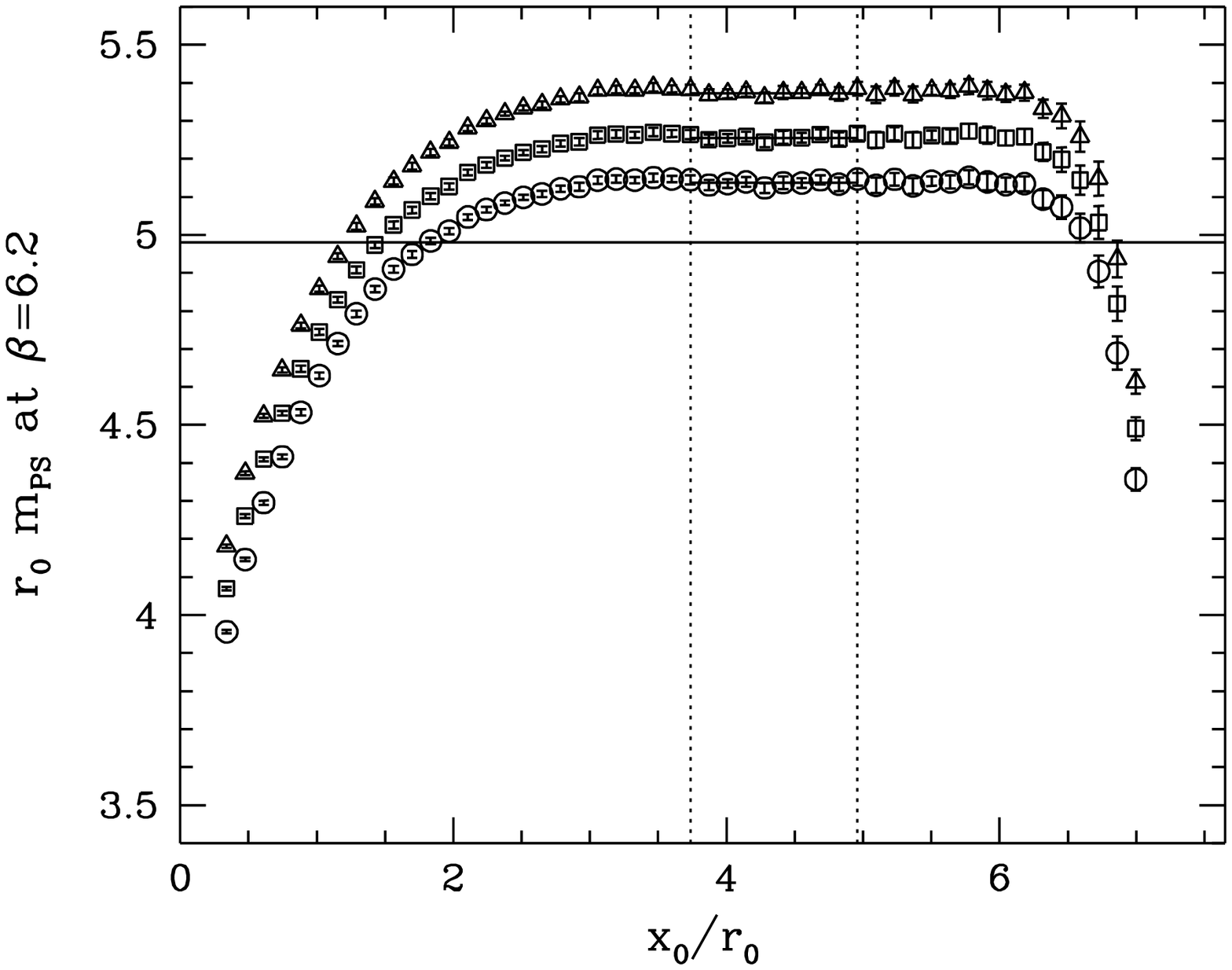, width=.48\linewidth,
        bbllx=18,bblly=164,bburx=572,bbury=608}
\epsfig{file=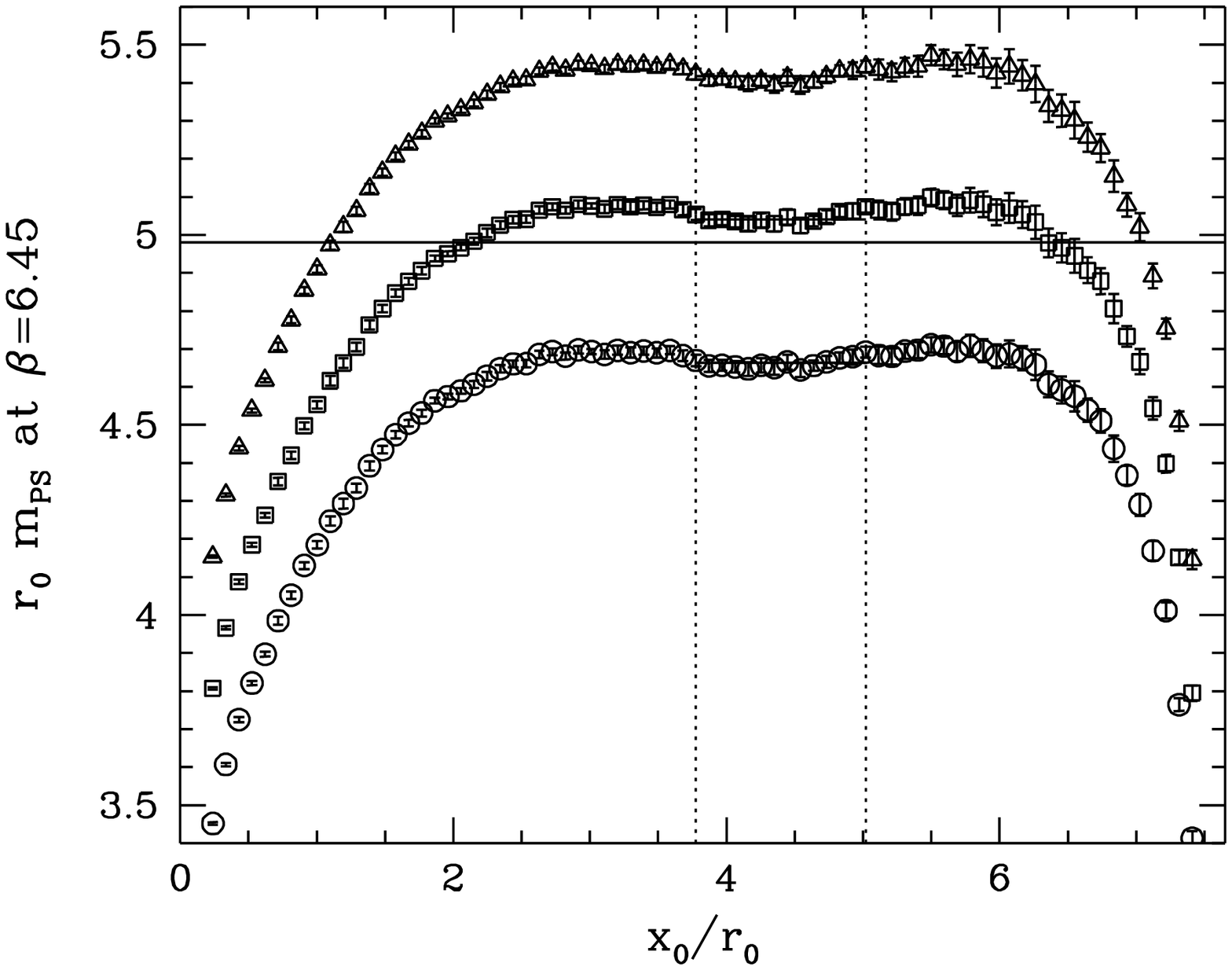, width=.48\linewidth,
        bbllx=18,bblly=164,bburx=572,bbury=608}\hfill
\end{minipage}
\caption{The effective mass plateaux from the correlation
function $f_{\rm A}(x_0)$ for 3 charm quark mass
parameters at all $\beta$ values. The plateaux regions 
are indicated by the vertical lines. The pseudoscalar meson masses
cover the physical $D_s$ meson mass (full line) except at $\beta=6.2$.}
\label{plateaux}
\end{figure}
\subsubsection{Excited states}

To estimate the contributions of excited states to the effective
masses we first chose an a priori range for the plateau.
Then we averaged $m_{\rm X}(x_0)$ over the plateau region and 
subtract the averaged value from $m_{\rm X}(x_0)$. 
For small $x_0$ we fit the remainder with an 
ansatz of the form $\eta \exp(-x_0 \Delta)$. Reasonable fits
could be obtained at all $\beta$ values, and could be used to 
quantify the contamination by excited states from the lower time
boundary. From the upper boundary at $x_0=T$ one expects a
contribution from the  $0^{++}$ glueball with mass 
$r_0m_{\rm G}\simeq 4.3$~\cite{Guagnelli:1999zf}. 
While in some cases a  rough confirmation seemed possible, 
we essentially assumed this to be the case.
In fact, given the problem with rounding errors
at large values of $x_0$, a clean signal for
the glueball state would have come as a surprise.

\subsubsection{Plateau regions}

Adding all the systematic errors linearly, a plateau region 
was determined as the interval of times $x_0$ where the
relative systematic error was smaller than some prescribed value.
We chose $0.3\%$ for the pseudoscalar mass, $0.5\%$
for the vector mass and $0.8\%$ for the mass splitting.
The plateau ranges found in this way were remarkably stable 
when expressed in physical units. Therefore we 
decided to define the final ranges as follows:
from $3.8\,r_0$ to $5.0\,r_0$ for the pseudoscalar masses, from $3.8\,r_0$ to
$5.2\,r_0$ for the vector masses and from $3.7\,r_0$ to $4.5\,r_0$ 
in the case of the mass splitting. 
At fixed $\beta$ and for fixed quark mass parameters, the meson masses
are then obtained as averages over the plateau region. Statistical
errors were determined by a jackknife procedure, and
the maximally allowed systematic error 
was taken as final systematic error on the effective masses.
This is likely to be an overestimate, as the systematic
error on the effective masses inside the plateaux is a bit smaller.
The results obtained in this way are displayed in table~\ref{tab_raw}.

\subsection{Determination of the charm quark mass}

The meson masses at the simulated parameter values,
can be considered as functions of the corresponding bare valence quark masses.
Besides the bare subtracted quark masses (obtained 
using the $\kappa_{\rm critical}$ values in table~\ref{tab_renorm}),
we used bare PCAC masses from the SF correlation functions
at $x_0=T/2$. An analysis of rounding errors for the
PCAC masses shows that these are at the level of $0.01\%$ and hence
completely negligible.
\begin{figure}[t]
\begin{minipage}[h]{\linewidth}
\epsfig{file=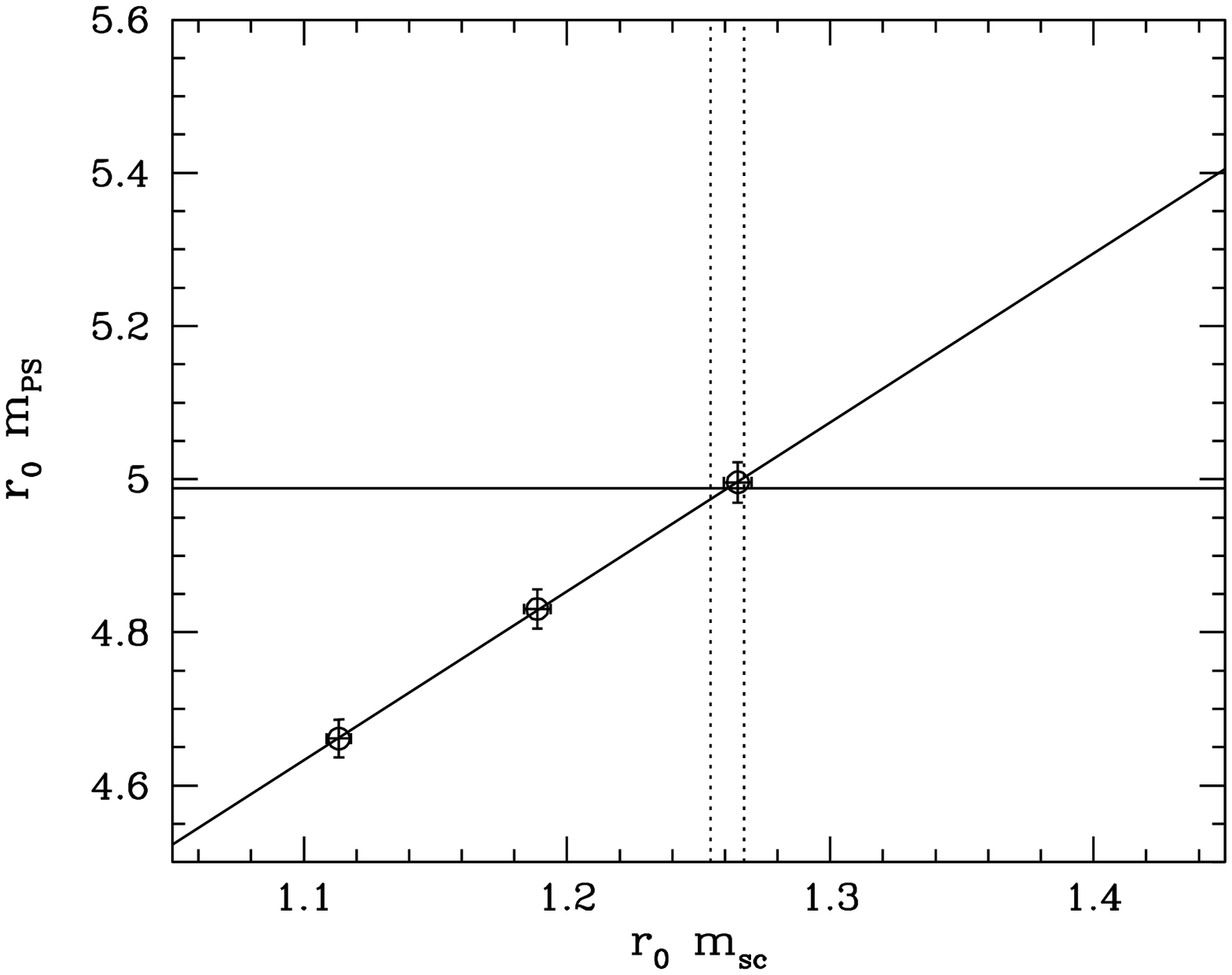, width=.48\linewidth,
        bbllx=18,bblly=144,bburx=572,bbury=518}
\epsfig{file=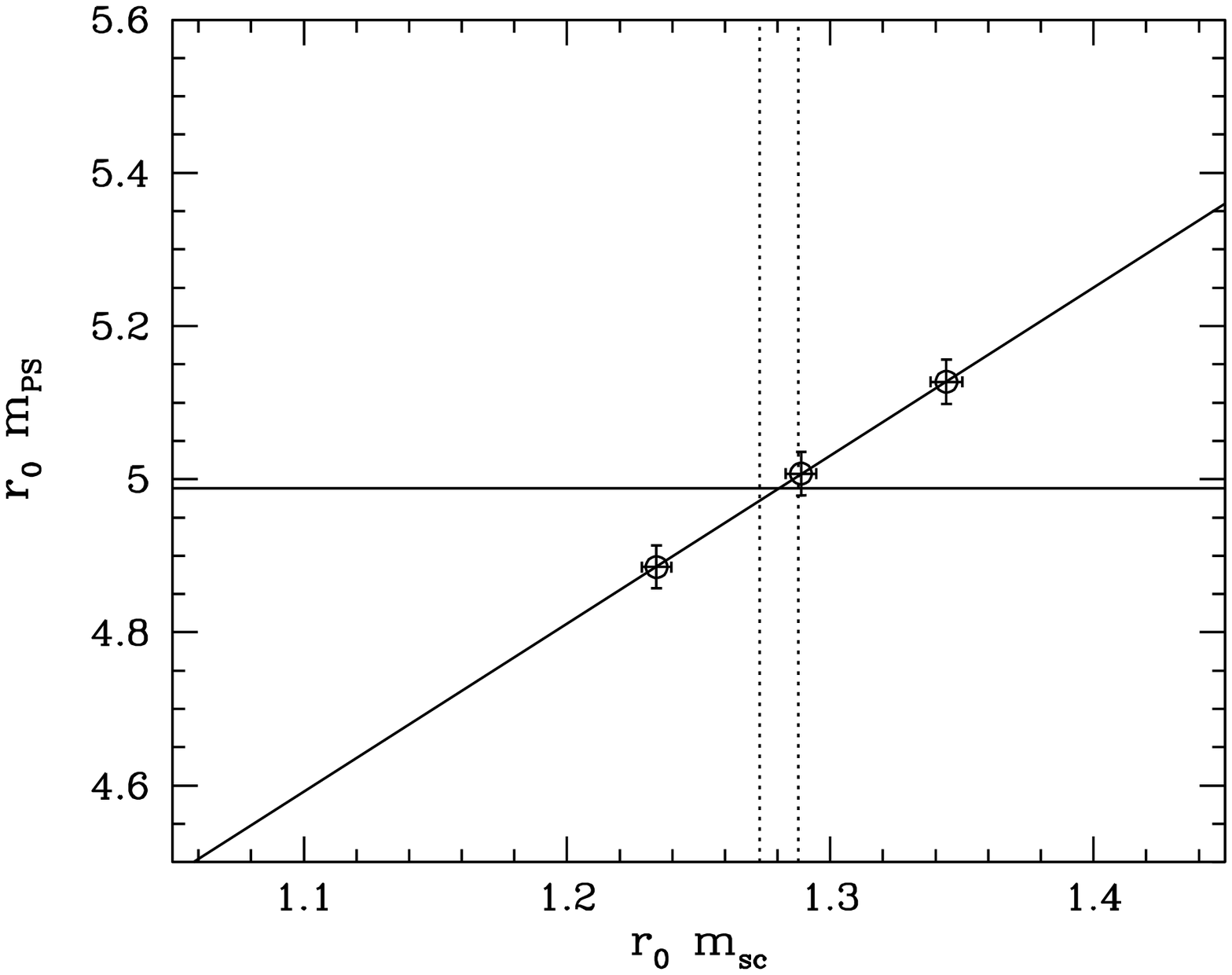, width=.48\linewidth,
        bbllx=18,bblly=144,bburx=572,bbury=518}\hfill\\
\epsfig{file=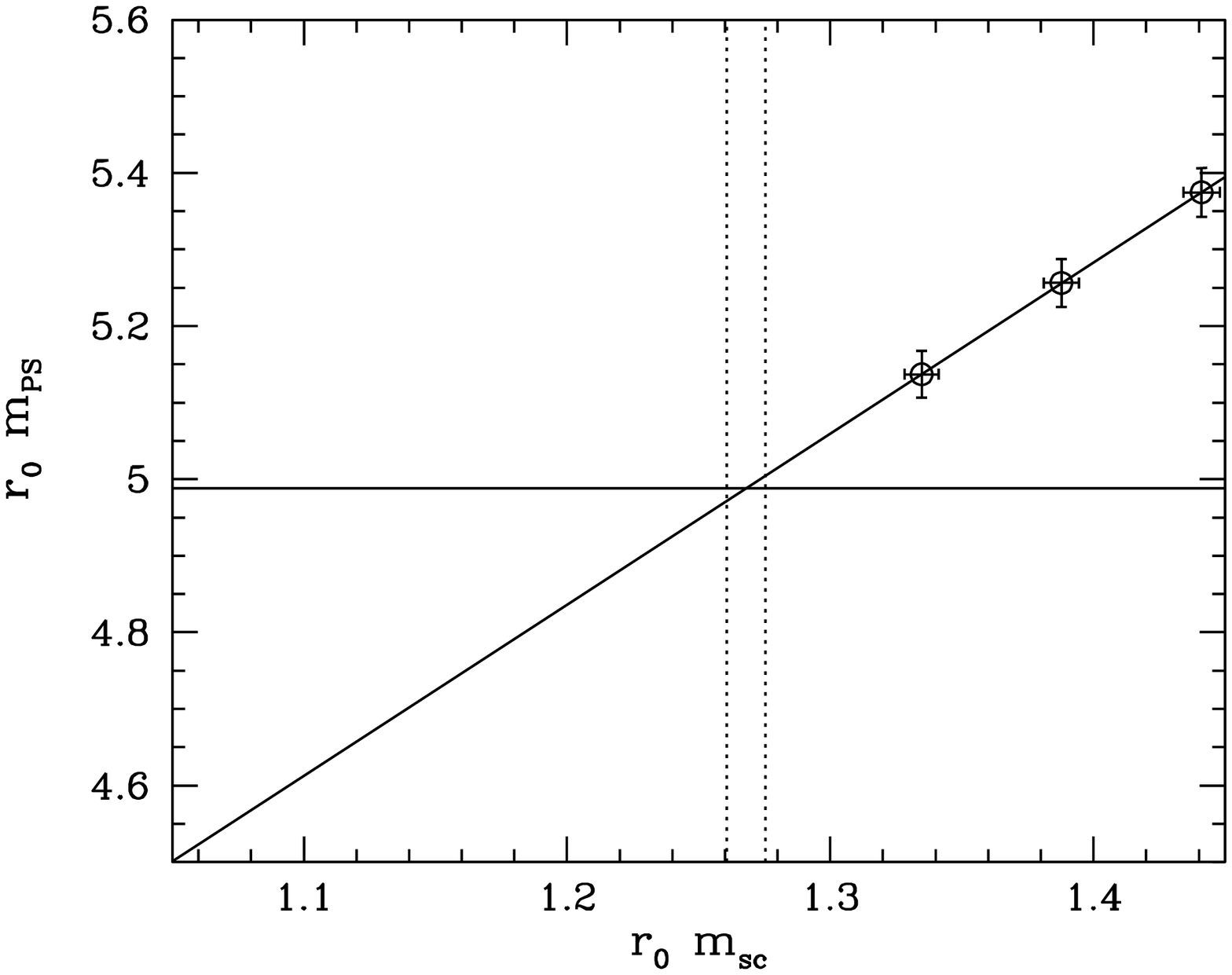, width=.48\linewidth,
        bbllx=18,bblly=164,bburx=572,bbury=608}
\epsfig{file=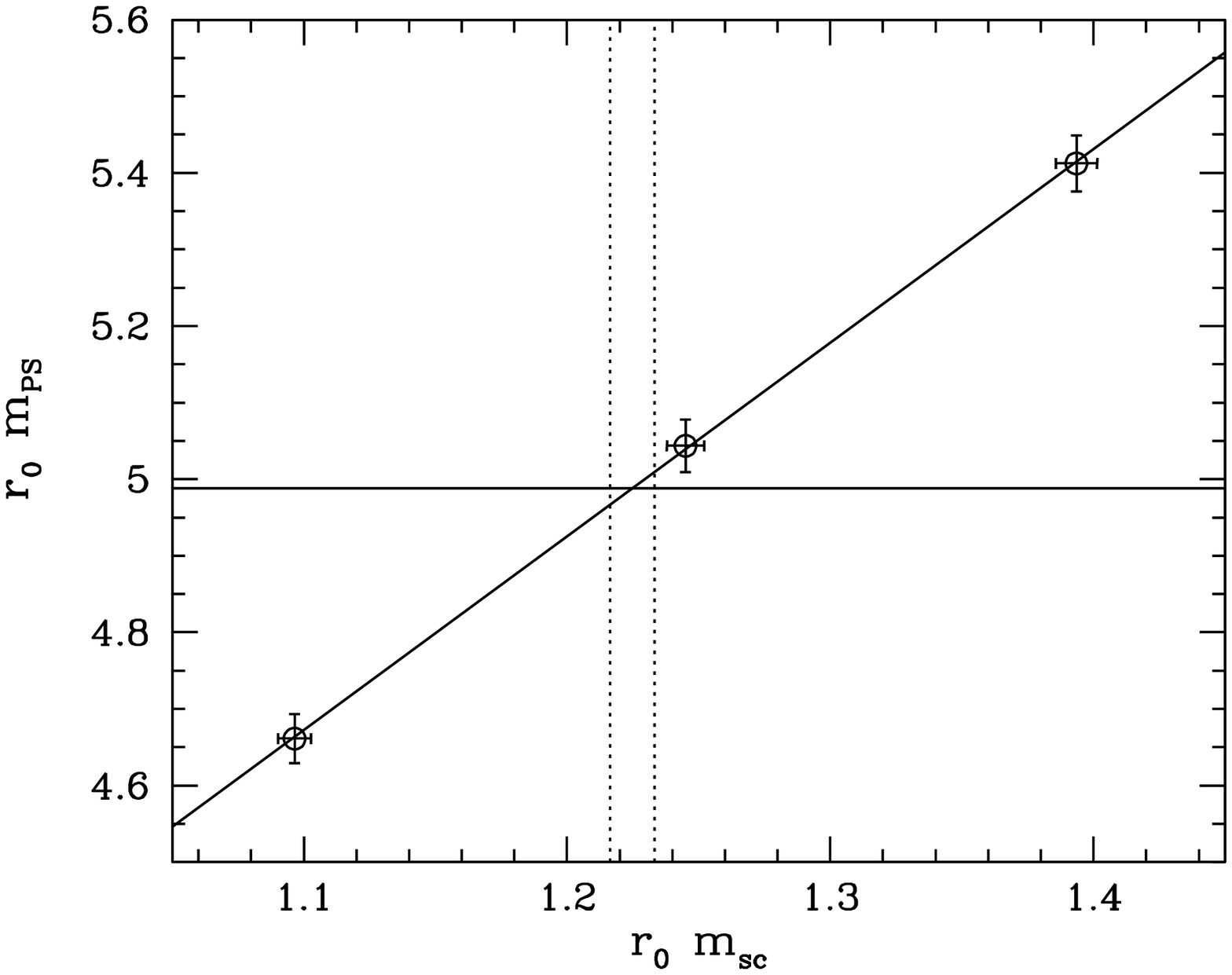, width=.48\linewidth,
        bbllx=18,bblly=164,bburx=572,bbury=608}\hfill
\end{minipage}
\caption{The inter- or extrapolation of $m_{\rm PS}$ as determined 
from figure~2 vs. the bare charm quark mass $r_0m_{\rm sc}$.}
\end{figure}
We distinguish the PCAC relation with
degenerate and non-degenerate quark masses. Furthermore,
we consider two choices of the lattice derivatives 
used in the PCAC relations: first, the standard choice
involving differences between nearest neighbours only,
and, secondly, the improved derivatives which also
involve next-to-nearest neighbours~[cf.~\cite{Guagnelli:2000jw} for
the definitions].
The results are given in tables~\ref{tab_sim1} and \ref{tab_raw}.

We recall that the strange quark mass parameters were already chosen
according to eq.~(\ref{m_strange}). Moreover, it turns
out that the meson masses only mildly depend on the strange
quark mass. Hence it remains to 
determine the bare charm quark masses where 
the pseudoscalar meson mass assumes its experimental
value, 
\begin{equation}
   r_0m_{\rm PS}=r_0m_{D_s} = 0.5\,\fm \times 1969\,\MeV = 4.99.
\end{equation}
Our simulation  parameters are such that this point 
could be reached by an interpolation, with 
the exception of $\beta=6.2$, where a small extrapolation was required. 
We performed linear fits of the form
\begin{equation}
   r_0m_{\rm PS}^{} = \alpha_0 + \alpha_1 r_0 m,
\end{equation}
where $m$ is some definition of the bare charm quark mass. 
The fits were done inside a Jackknife procedure to take
into account correlations of the data points.

\input tab_renorm

Given the various interpolated bare charm quark masses we 
construct the corresponding O($a$) improved RGI masses.
We distinguish 5 definitions, the first 3 being given by
\begin{eqnarray}
  r_0\Mc|_{\msc} & = & Z_M \Bigl\{2r_0 \msc 
  \left[ 1+(\ba-\bp)\tfrac12 (am_{\rm q,c}+am_{\rm q,s})\right] \nonumber\\
  &&\mbox{}-r_0m_{\rm s}\left[1+(\ba-\bp)am_{\rm q,s})\right]\Bigr\},
  \nonumber\\[1ex]
  r_0\Mc|_{\mc} &=&  Z_M r_0 \mc
  \left[ 1+(\ba-\bp)am_{\rm q,c}\right], \nonumber\\[1ex]
   r_0\Mc|_{m_{\rm q,c}} &=&  Z_M Z r_0 m_{\rm q,c}
   \left[ 1+\bm am_{\rm q,c}\right].
   \label{RGIs}
\end{eqnarray}
In addition we use the analogues of the first two 
definitions but with $\msc\rightarrow \msc^{\rm imp}$
and $\mc\rightarrow\mc^{\rm imp}$, i.e.~using
the next-to-nearest derivatives in the PCAC relations.
The results are shown in table~\ref{tab_results}, where the total
error is obtained by including the errors of $r_0/a$, 
and of the renormalization constants and improvement coefficients in 
quadrature~[cf.~table~\ref{tab_renorm}].

\input tab_results

%% file: tab_sim1.tex
\begin{table}[tb]
\centering
\begin{tabular}{l l c c c }
\hline \\[-1.0ex]
$\beta$ & $\kappa $ & $a\mq$ & $am$ & $am^{\rm imp}$ \\[1.0ex]
\hline \\[-1.0ex]
6.0  & 0.1190   & $0.5033$  & $0.5129(2)$ & $0.2962(3)$ \\
     & 0.1200   & $0.4683$  & $0.4736(2)$ & $0.2913(3)$ \\
     & 0.1210   & $0.4339$  & $0.4357(2)$ & $0.2837(2)$ \\[0.3ex]
     & 0.134108 & $0.0300$  & $0.0304(3)$ & $0.0299(4)$ \\
     & 0.133929 & $0.0350$  & $0.0353(3)$ & $0.0346(4)$ \\[1.0ex]
\hline \\[-1.0ex]
6.1  & 0.1218   & $0.4149$  & $0.4425(2)$ & $0.3113(2)$\\
     & 0.1224   & $0.3948$  & $0.4197(2)$ & $0.3030(2)$\\
     & 0.1230   & $0.3749$  & $0.3974(2)$ & $0.2940(2)$\\[0.3ex]
     & 0.134548 & $0.0260$  & $0.0317(3)$ & $0.0312(3)$\\
     & 0.134439 & $0.0290$  & $0.0348(3)$ & $0.0342(3)$\\[1.0ex]
\hline \\[-1.0ex]
6.2  & 0.1230   & $0.3830$  & $0.4072(2)$ & $0.3086(2)$ \\
     & 0.1235   & $0.3666$  & $0.3888(2)$ & $0.3000(2)$\\
     & 0.1240   & $0.3502$  & $0.3706(2)$ & $0.2909(2)$\\[0.3ex]
     & 0.134959 & $0.0228$  & $0.0254(2)$ & $0.0253(3)$ \\
     & 0.134832 & $0.0263$  & $0.0291(2)$ & $0.0288(2)$\\[1.0ex]
\hline \\[-1.0ex]
6.45 & 0.1270   & $0.2524$ & $0.2646(1)$ & $0.2336(1)$\\
     & 0.1280   & $0.2217$ & $0.2323(1)$ & $0.2094(1)$\\
     & 0.1290   & $0.1914$ & $0.2006(1)$ & $0.1844(1)$\\[0.3ex]
     & 0.135124 & $0.0157$ & $0.0166(1)$ & $0.0165(1)$ \\
\hline
\end{tabular}
\caption[]{Simulated $\kappa$-values and corresponding 
results for the bare subtracted and bare current
quark masses from the mass degenerate PCAC relation.}
\label{tab_sim1}
\end{table}

%% file: tab_renorm.tex
\begin{table}[tb]
\centering
\begin{tabular}{l c c c c  l}
\hline \\[-1.0ex]
$\beta$ & $\kappa_{\rm critical}$ & $Z_M$ & $Z$  &  $\ba-\bp$ & $\quad\bm$ \\[1.0ex]
\hline \\[-1.0ex]
6.0  & $0.135196$ & $1.752(19)$ & $1.0604(4)$ & $0.171(5)$ & $-0.709(6)$ 
\\[1.0ex]
6.1  & $0.135496$ & $1.782(20)$ & $1.0852(5)$ & $0.071(3)$ & $-0.699(5)$
\\[1.0ex]
6.2  & $0.135795$ & $1.807(20)$ & $1.0960(5)$ & $0.039(3)$ & $-0.691(7)$
\\[1.0ex]
6.45 & $0.135701$ & $1.852(20)$ & $1.1045(5)$ & $0.010(5)$ & $-0.673(12)$
\\[1.0ex]
\hline
\end{tabular}
\caption[]{Renormalization constants and improvement coefficients at
the simulated $\beta$-values.}
\label{tab_renorm}
\end{table}

%% file: tab_results.tex
\begin{table}[htb]
\centering
\begin{tabular}{l c c c c r r c c}
\hline \\[-1.0ex]
  $\beta$ & $r_0\Mc|_{\msc}$ & $r_0{\Mc}|_{\mc}$ 
  & $r_0\Mc|_{\mqc}$ & $r_0\Mc|_{\msc^{\rm im}}$ & $r_0\Mc|_{\mc^{\rm im}}$ 
  & $r_0m_{\rm V}^{}$ & $r_0m_{\rm S}^{}$ \\[1.0ex]
\hline \\[-1.0ex]
 6.0  & $4.331(59)$ & $5.215(75)$ & $3.224(41)$ & $3.566(47)$ & $3.026(37)$ & $5.281(14)$ & $0.299(11)$ \\
 6.1  & $4.274(59)$ & $4.824(70)$ & $3.479(43)$ & $3.755(50)$ & $3.492(44)$ & $5.299(15)$ & $0.303(12)$\\
 6.2  & $4.277(55)$ & $4.682(67)$ & $3.711(47)$ & $3.905(52)$ & $3.769(49)$ & $5.299(15)$ & $0.299(13)$\\
 6.45 & $4.220(60)$ & $4.428(64)$ & $3.975(53)$ & $4.038(57)$ & $3.995(56)$ & $5.300(22)$ & $0.300(17)$\\[1.0ex]
\hline\\[-1.0ex]
$CL$  & $4.19(11)$  & $4.20(12)$  & $4.27(10) $ & $4.21(11) $ & $4.31(11) $ & 
$5.300(35)$ & $0.297(26)$\\[1.0ex]
\hline
\end{tabular}
\caption[tab_param]{Results for the five definitions
of the RGI charm quark mass~[cf.~eqs.~(\ref{RGIs})], the vector meson
mass and the mass splitting between vector and pseudoscalar mesons.} 
\label{tab_results}
\end{table}

%% file: sect5.tex
\section{Continuum extrapolations and results}

\subsection{RGI charm quark masses}

We now come to our main results, the continuum extrapolation
of the RGI charm quark mass. 
As discussed in sect.~3, the charm quark mass can be
defined in various ways which are all equivalent up
to cutoff effects. Since complete O($a$) improvement has
been implemented, we attempt linear fits to the data of the form
\begin{equation}
    r_0M_{\rm c} =  A + B (a^2/r_0^2),
\end{equation}
with the two parameters $A$ and $B$.
As can be seen in fig.~4, the fits to the data appear very reasonable.
Excluding the coarsest lattice spacing, the $\chi^2$ per
degree of freedom ranges from $0.1$ to $1.5$. 
Moreover, the various definitions of $r_0M_{\rm c}$ all yield compatible
continuum extrapolated results. We also performed combined
fits, with the constraint of a common continuum limit. 
The correlation of the data at the same $\beta$-value 
was taken into account by determining the covariance matrix 
in a Jackknife procedure. However, the correlation is rather strong, so that
the error is not much reduced compared to the 
individual continuum extrapolations.

\begin{figure}[t]
\begin{minipage}[h]{\linewidth}
\epsfig{file=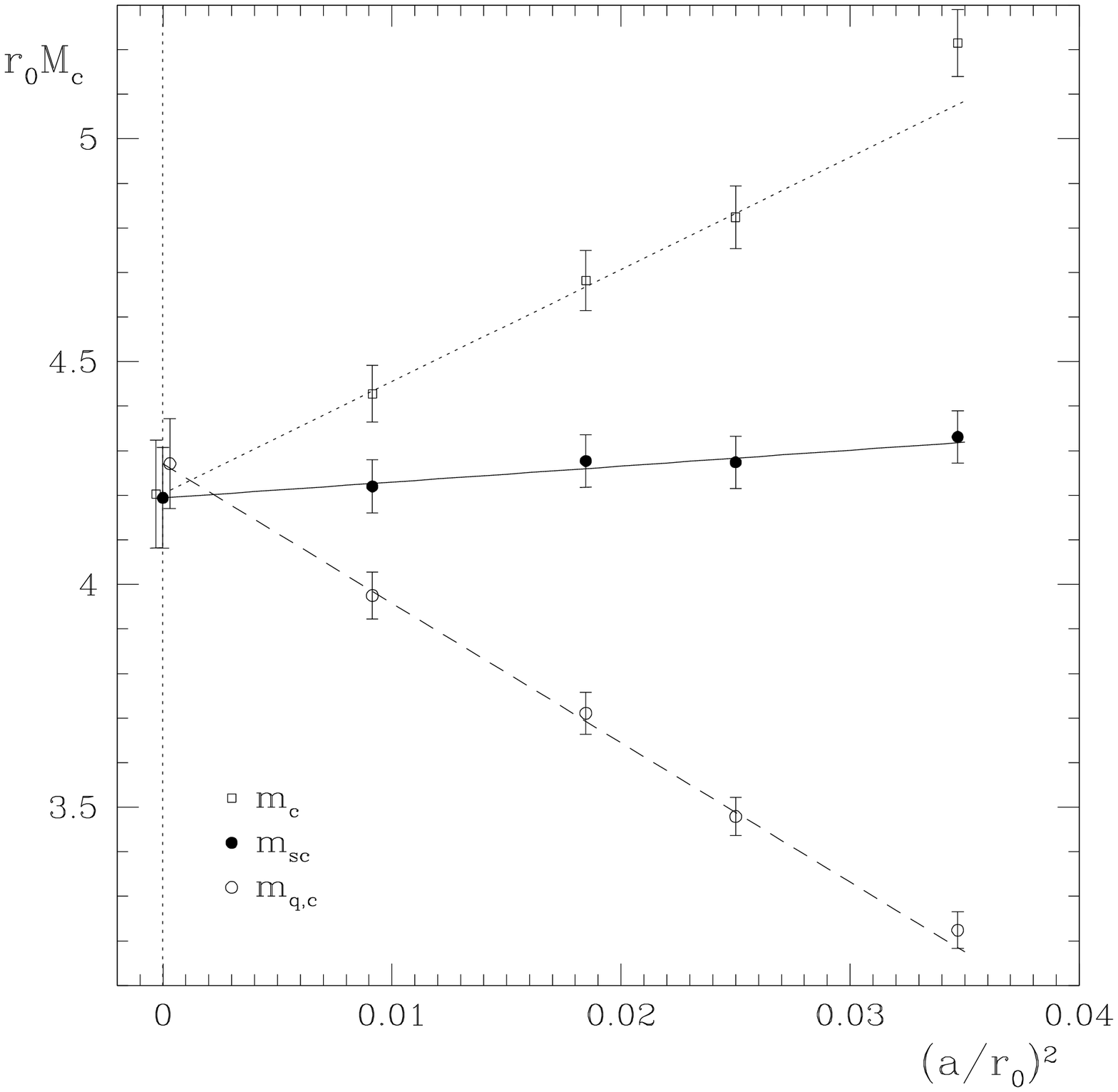, width=.48\linewidth}
\epsfig{file=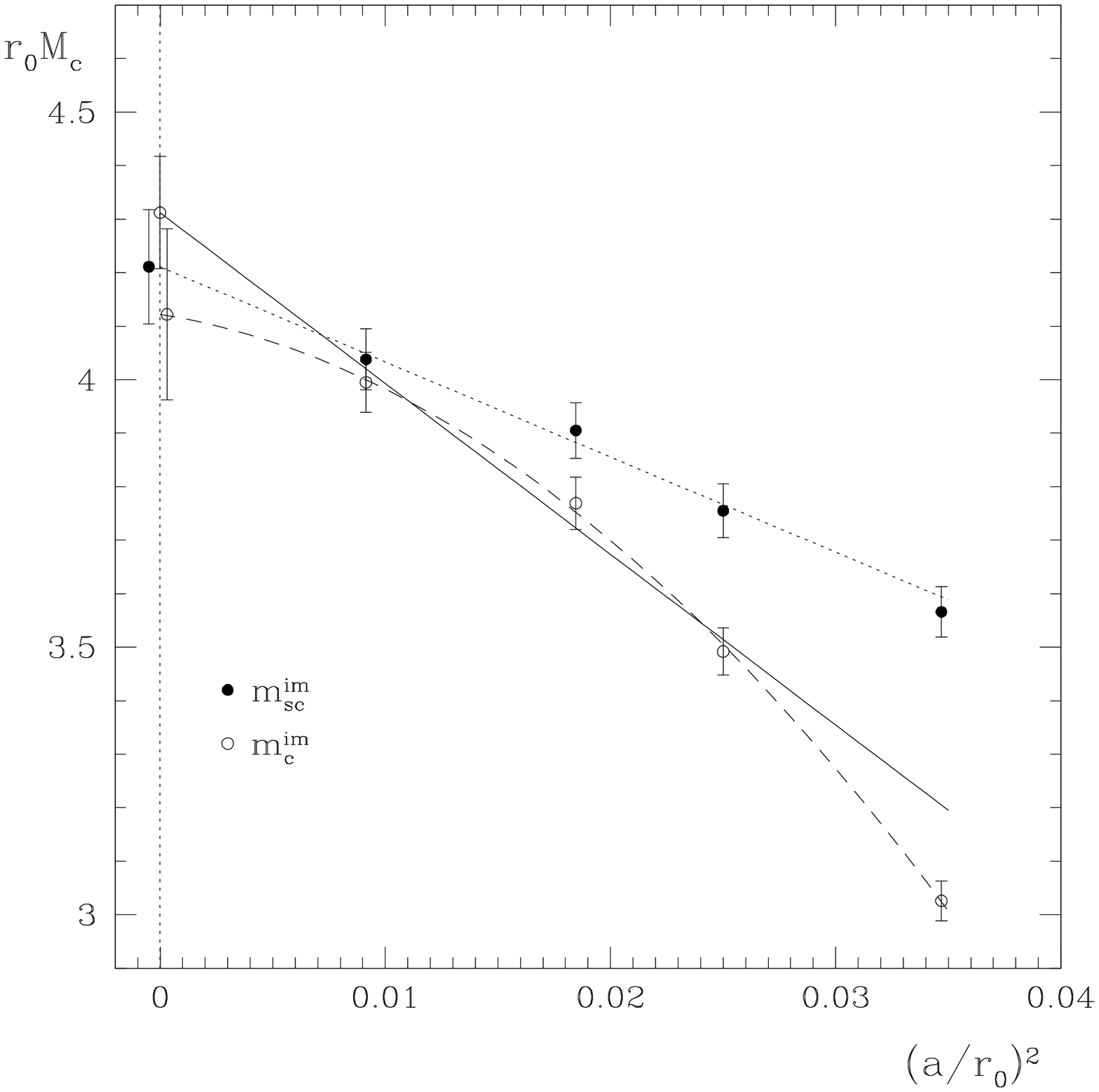, width=.48\linewidth}\hfill
\end{minipage}
\caption{The continuum extrapolation for all 5 definitions of  $r_0M_{\rm c}$ 
considered. In the case of $r_0\Mc|_{\mc^{\rm im}}$ higher order 
cutoff effects become visible. For illustration we also included a fit
to all 4 data points allowing for an additional term $\propto (a/r_0)^4$ (dashed line).} 
\end{figure}
As our best result we choose the one obtained from 
the heavy-light PCAC relation, as cutoff effects
are found to be rather small in this case. 
We quote
\begin{equation}
    r_0M_{\rm c} = 4.19(11),
  \label{r0Mc}
\end{equation}
where the error contains the $1.3$ percent error of the 
matching factor to the RGI mass~\cite{Garden:2000fg}.
In physical units we then have (with $r_0=0.5\,{\rm fm}$)
\begin{equation}
    M_{\rm c}= 1.654(45)\,{\rm GeV}.
\end{equation}

\subsection{Conversion to the $\MSbar$ scheme}

In our framework the RGI quark masses appear as
the primary quantities. However, many phenomenological applications 
use the masses in the $\MSbar$ scheme at the scale set by
the quark mass itself. In order to convert to this scheme, we
integrate the renormalization group equations
using the known 4-loop RG functions for 
$N_{\rm f}=0$~\cite{vanRitbergen:1997va,Chetyrkin:1997dh},
and find
\begin{equation}
    \mcmc = 1.301(28)(20)\,{\rm GeV} = 1.301(34)\,{\rm GeV},
  \label{mcmc}
\end{equation}
where we have used the result~(\ref{r0Lambda}) for $\Lambda_{\MSbar}^{(0)}$.
The two errors given in the first number correspond to 
the error in the mass $\Mc$ and 
the change induced by a variation of $\Lambda_{\MSbar}^{(0)}$ 
within its error bars respectively. 
As both errors are independent we have combined them 
quadratically in the last equation of~(\ref{mcmc}).
Note that the scale of the charm quark mass is already quite low,
so that the order of perturbation theory must be specified, too.
In fact, using 3 and 2-loop RG evolution we obtain, 
\begin{equation}
    \mcmc=\begin{cases}
                  1.294(34) \,\GeV & \text{3-loop evolution}, \\ 
                  1.263(34) \,\GeV & \text{2-loop evolution},
             \end{cases}
\end{equation}
i.e.~even the 4-loop contribution is still sizeable, and the
difference between 2-loop and 4-loop RG evolution is
as large as the total error.

\subsection{Further results}

\subsubsection{The mass ratio $\Mc/\Ms$}

A further consistency check for our results
is  provided by considering the ratio $\Mc/\Ms$.
Here one may compare the continuum extrapolation of 
the ratio $\Mc/\Ms$, to the ratio of the two continuum extrapolated
quark masses. Unfortunately, this exercise requires a rather 
precise tuning of the strange quark mass. 
As our strange quark mass parameters were tuned 
using interpolated values of $\kappa_{\rm critical}$, 
the corresponding curve $\kappa_{\rm s}(\beta)$ 
does not very precisely correspond to a condition 
of constant physics. 
Therefore, instead of using our own data for the strange quark mass, 
we decided to use the bare current quark masses from 
ref.~\cite{Garden:2000fg,Hartmut}, where a careful extrapolation
to the physical kaon mass has been performed. 
After O(a) improvement of the bare masses, we take the
ratios and combine the errors in quadrature.
The continuum extrapolation then yields
\begin{equation}
       \Mc/\Ms = 12.0(5),
\end{equation}
which can be compared to the ratio taken directly 
in the continuum limit, viz.
\begin{equation}
   \Mc/\Ms = 12.2(1.0).
\end{equation}

\subsubsection{$m_{D_s^\ast}$ and $m_{D_s}-m_{D_s^\ast}$}

We also computed the mass of the vector meson $D_s^\ast$. The
cutoff effects are very small, so that the continuum extrapolation
is not problematic. We obtain
\begin{equation}
       r_0m_{D_s^\ast} = 5.300(35) 
\qquad \Rightarrow \quad m_{D_s^\ast} = 2092(14) {\rm MeV}.
\label{vectormass}
\end{equation}
Despite the quenched approximation this is not far from the
experimental result, $m_{D_s^\ast} = 2112\, {\rm MeV}$~\cite{Groom:2000in}.
It is customary to study the mass splitting 
$m_{\rm S}= m_{D_s^\ast} - m_{D_s}$. Compared to the  experimental value
of $144\,{\rm MeV}$ the results of quenched lattice simulations
often turn out to be smaller (see, e.g.~\cite{Bowler:2000xw}). 
The mass splitting  may be obtained
in two ways: we may either subtract the input value $r_0m_{D_s}=4.99$
from the result (\ref{vectormass}), which yields
\begin{equation}
   r_0m_{D_s^\ast}-4.99 = 0.310(35) = 122(14)\,\MeV \times 0.5\,\fm.
\end{equation}
On the other hand, the mass splitting may be directly obtained
by studying the effective mass associated to the ratio of correlators
$f(x_0) = k_{\rm V}(x_0)/f_{\rm P}(x_0)$ (cf.~sect.~2). With this direct method
we obtain in the continuum limit
\begin{equation}
  r_0(m_{D_s^\ast}-m_{D_s}) = 0.297(26) = 117(11) \,\MeV \times 0.5\,\fm,
\end{equation}
which agrees with the previous result within errors.
Both continuum extrapolations are shown in figure~5,
and  we conclude that the mass splitting
is indeed smaller than the experimental value, albeit only by
1.5 and 2.5 standard deviations for the indirect and direct
methods respectively. 
\begin{figure}[h]
  \centering
  \epsfig{file=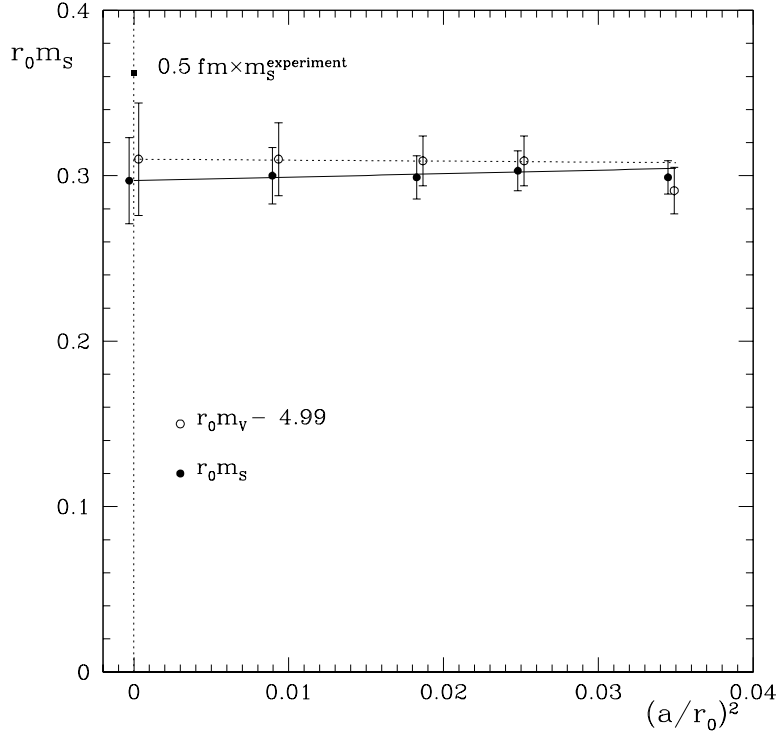, width=.7\linewidth}
  \caption{Results for the mass splitting $m_{\rm S}^{}= 
  m_{D_s^\ast}^{}-m_{D_s}$. The open points are obtained by
  computing the vector meson mass while the solid points represent
  the direct computation of $m_{\rm S}^{}$ from a ratio
  of correlation functions (the data points have been slightly
  offset for clarity).}
\end{figure}
\subsection{Quenched scale ambiguities}

As mentioned in sect.~2, the quenched approximation  
to QCD accounts for the observed hadronic spectrum up to 
inconsistencies which are at the 10 per cent level. 
Turning this around, the choice of different
hadronic input from experiment will lead to a spread of results
for the $\Lambda$ parameter and the quark masses in the
quenched approximation. Under the assumption
that these inconsistencies are entirely due to the  
neglected sea quark effects, this spread of results
may be taken as a first estimate of the quenching error.

We start with some definition of the RGI charm quark mass,
consider it as a function of $z=r_0 m_{D_s}$,
\begin{equation}
    r_0\Mc = F(z).
\end{equation}
and expand around the standard choice used so far,
$z_0=4.99$,
\begin{equation}
  F(z) = F(z_0) + (z-z_0) F'(z_0) +  \rmO\left((z-z_0)^2\right).
  \label{scale_shift}
\end{equation}
For a $10\%$ shift in the scale $r_0$ we
have $z-z_0 = 0.5$. With this choice we 
obtain an estimate of the first order term in the continuum limit,
\begin{equation}
   0.5 {F'}(4.99) = 0.7(1).
\end{equation}
Expressing $\Mc$ again in physical units (using now $r_0=0.55\, \fm$!)
we observe that the charm quark mass increases by about 6 percent.
Although this number is only considered a rough estimate we remark
that the approximations made are supported by our data:
first, figure~3 indicates that the higher order terms in 
eq.~(\ref{scale_shift}) are indeed small over quite some
range of charm quark masses. Second, we have
assumed that the shift in the strange quark mass can be neglected.
As observed in~\cite{Garden:2000fg}, the relative change in the 
light quark masses  is the same as for $r_0$ itself, 
due to the proportionality 
\begin{equation}
   M \propto r_0 m_{\rm PS}^2.
\end{equation}
As already mentioned earlier the sensitivity of the $D_s$ meson
mass to a variation of the strange quark mass is rather low,
and this can be seen in table~\ref{tab_raw}, where the differences
between the two chosen $\kappa_{\rm s}$-values correspond
to $10-15$ percent differences in the quark mass.

Having established the 6 percent increase of $\Mc$, 
we note that this corresponds to an increase of only 3 percent for 
$\mcmc$.
This is due to the quark mass anomalous 
dimension in the $\MSbar$ scheme, which is such that an increase of 
the renormalization scale decreases the running quark mass.
In passing we mention that this also holds for the strange
quark mass at the scale $\mu=2\, \GeV$ where the induced change
amounts to about 8 percent.

The same exercise can be done for the vector meson mass and
the mass splitting.  The first order terms are
now given by
\begin{equation}
   0.5 F'(4.99) = \begin{cases} 
                  \hphantom{+}
                   0.46(1) & \text{for $F = r_0m_{D_s^\ast}$}, \\[1ex]
                  -0.035(5)& \text{for $F = r_0(m_{D_s^\ast}-m_{D_s})$}.
                  \end{cases}
\end{equation}
When expressed in physical units this corresponds 
to changes of $+9\%$ and $-20\%$ respectively.
Hence, the quenched scale ambiguity for the mass splitting
is rather large. In particular, in this latter case  
the discrepancy to the experimental result increases 
to about 4 standard deviations. 

Finally we look at the ratio $\Mc/\Ms$, which may be expected
to be less sensitive to the scale ambiguity. 
Here, the $6\%$ shift of the charm quark mass
is overcompensated by the $10\%$ shift in
the strange quark mass, leaving a 
quenched scale ambiguity of about $4\%$.

\subsection{Taking $m_{D_s^\ast}$ as input}

Finally we consider setting the charm quark mass using 
the $D_s^\ast$-meson instead of $D_s$ as input. The analysis 
is completely analogous and the continuum extrapolations
look qualitatively very similar. The $D_s$ mass becomes
now a measurement, and the result,
\begin{equation}
   r_0D_s = 5.04(4),
\end{equation}
is not far from the experimental value $4.99$ previously used as input.
As our best result for the charm quark mass we
quote again the value obtained from the mass non-degenerate
PCAC mass. The result,
\begin{equation}
   r_0M_{\rm c} = 4.27(13),
\end{equation}
is slightly larger but 
consistent within errors with eq.~(\ref{r0Mc}).

%% file: sect6.tex
\section{Conclusions}

The main result of this paper is the determination of 
the RGI charm quark mass using  
the experimentally measured $D_s$ meson mass as essential
input. Apart from the quenched approximation
all errors appear well under control.  With specified
input the total error is around 3 percent,
which is smaller than the error one might attribute
to the use of the quenched approximation. An
estimate of the latter has been obtained by varying the
scale which is used to assign physical units;
assuming that the real world is described by full QCD, 
the inconsistencies may be taken as a first indication 
of the quenching error. Surprisingly, a scale variation
of $10\%$ induces only a $6\%$ shift of the RGI charm quark mass,
which is further reduced to $3\%$ for the charm quark mass in 
the $\MSbar$ scheme at the charm quark mass scale.

While the agreement with some non-lattice charm quark mass
determinations is surprisingly good~(in particular
the value $\mcmc=1304(27)\,\MeV$ of ref.~\cite{Kuhn:2001dm})
we emphasize that the real quenching error can only be asserted by 
going beyond the quenched approximation.
However, our results show the way towards more realistic studies, 
as well as the potential strength of lattice techniques.
In particular,  we draw the following conclusions:
\begin{itemize}
\item
  precise results in charm physics are attainable using
  the standard set-up of O($a$) improved lattice QCD,
  which was originally designed for the light quarks, 
\item
  cutoff effects can be quite large in general, making
  a continuum extrapolation necessary. In our examples
  decent continuum extrapolations were possible
  based on data covering a factor of 2 in the cutoff scale.
\end{itemize}
Finally, as a by-product of our simulations we measured
the $D_s^\ast$ meson mass and the mass splitting between
the vector and pseudoscalar $D_s$ mesons. The latter turns
out to be rather small when compared to experiment, 
and is very sensitive to the quenched 
scale ambiguity. It will be interesting to look at the
effect of sea quarks on the mass splitting.

%% file: acknow.tex
\vskip 1ex
This work is part of the ALPHA collaboration research programme,
and partially supported by the European Community under the 
grant HPRN-CT-2000-00145 Hadrons/Lattice QCD.
Simulations were carried out on machines of the APE100 and APE1000
series at DESY-Zeuthen. We thank the staff at the computer centre
for their help, and P.~Ball, B.~Bunk, T.~Hurth, R.~Sommer,
H.~Wittig and U.~Wolff for useful discussions.

\vfill
\eject

%% file: appa.tex
\appendix
\renewcommand{\thesection}{A}
\section{Raw lattice data}
\renewcommand{\thesection}{A}

\input tab_rawres.tex

\vfill\eject

%% file: tab_rawres.tex
\begin{table}[htb]
\centering
\begin{tabular}{c c c c c c c c}
\hline \\[-1.0ex]
$\beta$ & $\kappa_{\rm c}$ & $\kappa_{\rm s}$ & $a\msc$ & $a\msc^{\rm imp} 
$ & $am_{\rm PS}^{}$ & $am_{\rm V}^{}$ 
& $am_{\rm S}^{}\!\times\! 10^2$ \\[1.0ex]
\hline \\[-1.0ex]
6.0  & 0.1190 & 0.134108 & 0.2356(3) & 0.1962(4) & 0.9306(16) & 0.985(3) & 5.55(20)\\
     &        & 0.133929 & 0.2387(3) & 0.1982(4) & 0.9366(15) & 0.990(3) & 5.43(20)\\[0.3ex]
     & 0.1200 & 0.134108 & 0.2214(3) & 0.1869(4) & 0.8998(16) & 0.957(3) & 5.80(20)\\
     &        & 0.133929 & 0.2245(3) & 0.1890(3) & 0.9058(15) & 0.962(3) & 5.67(20)\\[0.3ex]
     & 0.1210 & 0.134108 & 0.2074(3) & 0.1775(3) & 0.8683(15) & 0.928(3) & 6.06(21)\\
     &        & 0.133929 & 0.2104(3) & 0.1796(4) & 0.8744(15) & 0.933(3) & 5.93(20)\\[1.0ex]
\hline \\[-1.0ex]
6.1  & 0.1218 & 0.134548 & 0.2125(3)  & 0.1871(3) & 0.811(2) & 0.858(3)  & 4.68(21)\\
     &        & 0.134439 & 0.2144(3)  & 0.1885(3) & 0.815(2) & 0.862(3)  & 4.64(20)\\[0.3ex]
     & 0.1224 & 0.134548 & 0.2038(3)  & 0.1806(3) & 0.792(2) & 0.841(3)  & 4.81(21)\\
     &        & 0.134439 & 0.2057(3)  & 0.1820(3) & 0.796(2) & 0.844(3)  & 4.77(20)\\[0.3ex]
     & 0.1230 & 0.134548 & 0.1951(3)  & 0.1740(3) & 0.772(2) & 0.823(3)  & 4.96(21)\\
     &        & 0.134439 & 0.1970(3)  & 0.1754(3) & 0.776(2) & 0.826(3)  & 4.92(20)\\[1.0ex]
\hline \\[-1.0ex]
6.2  & 0.1230 & 0.134959 & 0.1958(2)  & 0.1770(2) & 0.730(2) & 0.769(3)  & 3.73(19)\\
     &        & 0.134832 & 0.1980(2)  & 0.1787(3) & 0.735(1) & 0.773(2)  & 3.67(17)\\[0.3ex]
     & 0.1235 & 0.134959 & 0.1886(2)  & 0.1713(2) & 0.714(2) & 0.754(3)  & 3.84(18)\\
     &        & 0.134832 & 0.1907(2)  & 0.1730(2) & 0.719(1) & 0.758(2)  & 3.77(17)\\[0.3ex]
     & 0.1240 & 0.134959 & 0.1814(2)  & 0.1655(3) & 0.698(1) & 0.739(3)  & 3.95(19)\\
     &        & 0.134832 & 0.1835(2)  & 0.1673(3) & 0.703(1) & 0.743(2)  & 3.88(17)\\[1.0ex]
\hline \\[-1.0ex]
6.45 & 0.1270 & 0.135124 & 0.1332(2)  & 0.1270(2) & 0.518(2) & 0.544(3)  & 2.59(18)\\
     & 0.1280 &          & 0.1191(2)  & 0.1142(2) & 0.482(1) & 0.512(3)  & 2.82(20)\\
     & 0.1290 &          & 0.1049(2)  & 0.1012(2) & 0.446(1) & 0.478(3)  & 3.09(20)\\
\hline
\end{tabular}
\caption[]{Results for unrenormalized current quark and meson masses
for all heavy-light combinations of simulated $\kappa$-values (errors are
statistical only).}
\label{tab_raw}
\end{table}